\documentclass[aps,  % American Physical Society (APS) styling
prc,                 % customize styling for APS journals
twocolumn,           % two-column formatting
floatfix,            % avoid float problems when twocolumn is used
nofootinbib,
longbibliography
groupedaddress]
{revtex4-2}

\usepackage{amsmath}
\usepackage{bm}
\usepackage{graphicx}
\usepackage{color}
\usepackage[colorlinks=true, linkcolor=black, citecolor=blue, urlcolor=blue]{hyperref} 
\usepackage[english]{babel}
\usepackage[ansinew]{inputenc}
\usepackage[usenames,dvipsnames]{xcolor}
\usepackage{overpic}
\usepackage{soul,xcolor}

\def\x{\mathbf{x}}
\begin{document}

\title{Nonequilibrium effects and transverse spherocity in ultra-relativistic proton-nucleus collisions}

\author{Lucia Oliva}
\email{lucia.oliva@dfa.unict.it}
\affiliation{Institute for Theoretical Physics, Johann Wolfgang Goethe Universit\"{a}t, Frankfurt am Main, Germany}
\affiliation{Department of Physics and Astronomy ``Ettore Majorana'', University of Catania,
Via S. Sofia 64, I-95123 Catania, Italy}
\affiliation{INFN Sezione di Catania, Via S. Sofia 64, I-95123 Catania, Italy}

\author{Wenkai Fan}
\affiliation{Department of Physics, Duke University, Durham, NC 27708, USA}

\author{Pierre Moreau}
\affiliation{Department of Physics, Duke University, Durham, NC 27708, USA}

\author{Steffen A. Bass}
\affiliation{Department of Physics, Duke University, Durham, NC 27708, USA}

\author{Elena Bratkovskaya}
\affiliation{Institute for Theoretical Physics, Johann Wolfgang Goethe Universit\"{a}t, Frankfurt am Main, Germany}
\affiliation{GSI Helmholtzzentrum f\"{u}r Schwerionenforschung GmbH, Planckstrasse 1, 64291 Darmstadt, Germany}
\affiliation{Helmholtz Research Academy Hesse for FAIR (HFHF), GSI Helmholtz Center for Heavy Ion Physics, Campus Frankfurt, 60438 Frankfurt, Germany}

%%%%%%%%%%%%%%%%%%%% Abstract %%%%%%%%%%%%%%%%%%%%%

\begin{abstract}
We investigate the effects of nonequilibrium dynamics in small colliding systems by comparing a nonequilibrium transport approach, the Parton-Hadron-String-Dynamics (PHSD), with a (2+1)D viscous hydrodynamic model, VISHNew. Focusing on p+Pb collisions at LHC energy, we extract the initial conditions for the hydrodynamic model from PHSD, in order to reduce the impact of the early out-of-equilibrium dynamics and focus on the traces of nonequilibiurm in the ensuing medium evolution. We find that in the transport approach quantities like energy density and bulk viscous pressure are highly inhomogeneous on the transverse plane during the whole evolution, whereas the hydrodynamic simulations dissolve more efficiently the initial spatial irregularities, still keeping a high degree of inhomogeneity due to the smaller size and lifetime of the medium produced in p+Pb collisions with respect to heavy-ion reactions.
As a first step that will help to identify the impact of these nonequilibrium effects on final observables in proton-nucleus collisions, we perform an analysis of the transverse spherocity, an event-shape observable able to distinguish between jetty and isotropic configurations of transverse momenta. We found that the spherocity distribution in PHSD is slightly shifted towards the isotropic limit with respect to the hydrodynamic result. Even though this dissimilarity is partially due to the difference in the final charged particle production, it mainly comes from the different description within the two frameworks of the medium produced in small colliding systems. This finding supports the idea that multi-differential measurements, such as those based on event categorization according to multiplicity and spherocity, are useful to study final-state observables in ultrarelativistic proton-nucleus collisions.
\end{abstract}

\pacs{}
\keywords{}
\maketitle

%----------------------------------------------------------------
\section{Introduction}
\label{sec:intro}

The high-energy collisions performed at the Relativistic Heavy Ion Collider (RHIC) and at the Large Hadron Collider (LHC) aim to have experimental access to hot QCD matter, since a deconfined state of quarks and gluons is produced thanks to the extreme condition of temperature achieved in such collisions. Signatures of the presence of the Quark-Gluon Plasma (QGP) were previously considered exclusive of heavy-ion experiments, while the measurements in high-multiplicity proton-proton ($pp$) and proton-nucleus ($pA$) collisions have recently shown similar features that could be explained with the formation of a strongly interacting medium \cite{Nagle:2018nvi}. 
Among those signatures, collective flow behaviour \cite{ALICE:2014dwt, ATLAS:2017rtr, CMS:2017kcs, PHENIX:2018lia} and the enhanced production of strange hadrons \cite{STAR:2007cqw, ALICE:2016fzo} are indicative of modifications of the system evolution and hadronization due to the presence of a dense medium.
The observed QGP signals in small systems raise the issue of thermal and chemical equilibration in such collisions where the possibly formed QGP should be produced as short-lived droplets, i.e., with a smaller space-time size. In this respect the understanding of the nonequilibrium effects in small colliding system acquires further importance.
The investigation of such out-of-equilibrium phenomena can be efficiently pursued by comparing two models, that have been the pillars of the medium descriptions in relativistic heavy-ion collisions and are now commonly used also for the study of high-multiplicity $pp$ and $pA$ collisions: the microscopic transport approach and the macroscopic hydrodynamic formulation. 
Even though, traditionally, an approximate local equilibrium was considered a necessary
condition for the validity of viscous hydrodynamics, the recent success in reproducing experimental data for small systems -- where the medium is probably quite far from local equilibrium -- has led to a change of paradigm from equilibration to hydrodynamization (i.e. the onset of the regime where hydrodynamics is applicable) \cite{Romatschke:2016hle}.
This new paradigm has in fact raised a lot of attention on studies on hydrodynamic attractors \cite{Heller:2015dha} for understanding the onset of fluid-dynamic behavior.
The transport approach is inherently a nonequilibrium dynamical model, hence it treats in a suitable way the out-of-equilibrium processes like those affecting the system in the very early stages of relativistic collisions.
A goal of this paper is the study of the evolution of the medium produced in proton-nucleus collisions and its nonequilibrium traces, by comparing a transport and a viscous hydrodynamic description, which have fundamentally different dynamics and treat in a different way the nonequilibrium effects present in the two approaches.
To this end we perform simulations of p+Pb collisions at LHC energy of $\sqrt{s_{NN}}=5.02$ TeV with the microscopic Parton-Hadron-String Dynamics (PHSD) transport approach  \cite{Cassing:2008sv,Cassing:2009vt,Cassing:2008nn} and the VISHNew hydrodynamic model \cite{Song:2007ux,Song:2008si,Shen:2014vra}.
In order to reduce the impact of the early out-of-equilibrium stage and focusing on the later evolution of the medium, the initial conditions for the hydrodynamic model are extracted from PHSD, which describes the full space-time evolution of the relativistic nuclear collision from the initial hard scatterings and string formation through the dynamical onset of the deconfined QGP phase to the hadronization and subsequent interactions in the hadronic phase.
A similar strategy has been used in Refs.~\cite{Xu:2017pna, Moreau:2017hxh} for analysing macroscopic properties of the QGP medium formed in relativistic heavy-ion collisions, such as thermodynamic quantities and viscous corrections as well as spatial and momentum eccentricities. The comparison has been extended in Ref.~\cite{Song:2020tfm} to heavy quark interactions with the expanding QGP modeled by means hydrodynamics or PHSD, in order to identify the effects of nonequilibrium matter on charm-quark dynamics in heavy-ion collisions.

The difficulty to well identify QGP signals in small systems has lead to first attempts to study observables connected to particle production and flow-like phenomena through novel multi-differential methods.
Transverse spherocity, originally proposed in Ref.~\cite{Banfi:2010xy}, is an event-shape observable, able to distinguish events according to their ``jetty'' or ``isotropic'' topology.
The usefulness of event-shape analysis with spherocity has been discussed for p+p collisions in few theoretical \cite{Khuntia:2018qox} and experimental studies \cite{Acharya:2019mzb, Nassirpour:2020owz}. 
Refs. \cite{Prasad:2021bdq, Mallick:2020ium, Mallick:2020dzv} address this issue with theoretical models in the context of heavy-ion collisions.
In this paper we perform a first analysis of event-shape engineering with spherocity in proton-nucleus collisions, investigating also the difference coming from the medium evolution description.
This is obtained by simulating the system produced in p+Pb collisions at LHC energy by means of PHSD and a hybrid model that includes VISHNew, the latter being initializated with thermodynamic quantities extracted from PHSD. In this way we focus on the evolution of the approximately thermalized medium, still keeping the nonequilibrium information that are intrinsically included in the tranport PHSD description.
Event categorization according to transverse spherocity in relativistic nuclear collisions may flank the current event-shape engineering based on flow vector \cite{Poskanzer:1998yz} addressed especially in LHC experiments for heavy-ion collisions \cite{ATLAS:2015qwl}.

%---------------------------------------------------------------
\section{Model description and initialization}
\label{sec:models}

The dynamical evolution of heavy-ion collisions as well as small colliding systems at relativistic energy, such as the p+Pb collisions at LHC energy that are the focus of this paper, can be described mainly within two different models: the microscopic transport approach or the macroscopic hydrodynamic formulation.
The former is represented in this work by the PHSD approach, whereas the VISHNew model is used for the latter description.
The two models are described in the next subsections together with the explanation of the initialization of the hydrodynamic code by means of initial conditions extracted from the early evolution of the medium in PHSD.

\subsection{Model I: PHSD transport approach}
\label{sec:phsd}

The Parton-Hadron-String Dynamics (PHSD) approach is a covariant dynamical model for strongly interacting many-body systems based on generalized transport equations, which are derived from the off-shell Kadanoff-Baym equations for nonequilibrium Green functions in phase-space representation \cite{Juchem:2003bi, Juchem:2004cs, Cassing:2009vt, Cassing:2008sv, Cassing:2008nn}.
In the Kadanoff-Baym theory the field quanta are treated as dressed propagators with complex self-energies, whose real and imaginary parts are related respectively to mean-field potentials and particle widths \cite{Cassing:2008nn}.
The off-shell transport equations fully describe the time evolution of the many-body system both in the partonic and in the hadronic phase, once the complex self-energies of the proper degrees of freedom are known \cite{Juchem:2003bi, Juchem:2004cs, Cassing:2008nn}.

PHSD simulates the full space-time evolution of the collision since the primary nucleon-nucleon inelastic scatterings between the two impinging nuclei. These lead to the formation of color-neutral strings described by the FRITIOF model \cite{Andersson:1992iq} based on the Lund string fragmentation picture.
The strings fragment into ``pre-hadrons'', i.e. baryons and mesons within their formation time ($\tau_f\approx0.8$ fm/$c$ in their rest frame) that do not interact with the surrounding medium, and into ``leading hadrons'', i.e. the fastest residues of the string ends that can reinteract with other hadrons with reduced cross sections.
If the local energy density $e$ is above the critical value of the deconfinement transition ($e_c\approx0.5$ GeV/fm$^3$), pre-hadrons dissolve in massive quarks, antiquarks and gluons plus a mean-field potential.

The Dynamical Quasi-Particle Model (DQPM) \cite{Cassing:2009vt} describes the properties of the QGP defining the parton spectral functions, i.e. masses $M_{q,g}(e)$ and widths $\Gamma_{q,g}(e)$, and self-generated repulsive mean-field potentials $U_{q,g}(e)$. Within the DQPM $e$ is related through the lattice QCD Equation of State (EoS) \cite{Borsanyi:2012cr, Borsanyi:2013bia} to the local temperature $T$.
In the DQPM the temperature-dependent effective masses and widths of quasi-particles are fitted to the lattice QCD thermodynamic quantities, such as energy density, pressure and entropy density.
In Ref.~\cite{Moreau:2019vhw} PHSD has been extended by including for all binary scattering channels in the partonic phase the dependence not only on the temperature but also on the baryon chemical potential $\mu_B$. Nevertheless, in this work we have used the default version due to the high collision energy considered.
The transport properties of the QGP, such as shear and bulk viscosities as well as electric conductivity, can be determined from the partonic interaction rates derived from the DQPM and result in line with the corresponding coefficients computed on the lattice \cite{Ozvenchuk:2012kh, Cassing:2013iz, Soloveva:2019xph}.
The transition from the partonic to hadronic degrees of freedom is described by dynamical hadronization with covariant transition rates for the fusion of quark-antiquark pairs to mesonic resonances and three quarks or antiquarks to baryonic states \cite{Cassing:2009vt, Bratkovskaya:2011wp}.
Due to the off-shell nature of both partons and hadrons, the hadronization process fulfils energy-momentum conservation, flavor-current conservation and color neutrality; moreover, it obeys the second law of thermodynamics of total entropy increase.

In the hadronic phase, i.e., for energy densities below $e_c$, the PHSD approach is equivalent to the Hadron-Strings Dynamics (HSD) model \cite{Cassing:1999es}.
PHSD has been widely used to simulate nucleus-nucleus collisions from the lower superproton-synchrotron (SPS) to the top LHC energies, obtaining a good description of bulk observables, collective flows and hard and electromagnetic probes \cite{Cassing:2009vt, Bratkovskaya:2011wp, Konchakovski:2011qa, Konchakovski:2012yg, Konchakovski:2014fya, Linnyk:2015rco}. Furthermore, PHSD has been applied to proton-nucleus collisions at RHIC and LHC energies for the study of final particles distribution and collective flows \cite{Konchakovski:2014wqa, Oliva:2019kin}.

\subsection{Model II: hydrodynamic approach + hadronic afterburner}
\label{sec:Duke}

The second model is a hybrid approach that simulates the hot and dense QGP phase of the nuclear collision using relativistic viscous hydrodynamics and the cooler and more dilute regions of the fireball employing relativistic Boltzmann transport equations \cite{Moreland:2018gsh, Moreland:2019szz}. The two phases are separated by a switching temperature $T_{switch}$: the medium above this temperature is simulated hydrodynamically and matter below $T_{switch}$ is described with a microscopic transport model.
The partonic phase is simulated with VISHNew, that is a boost-invariant viscous hydrodynamics model in 2+1 dimensions \cite{osuhydro}. VISHNew is an upgrade of VISH2+1 \cite{Song:2007ux, Song:2008si} that is able to handle fluctuating event-by-event initial conditions \cite{Shen:2014vra}.
The code has been extensively validated and reproduces semianalytic solutions of ideal hydrodynamics with an excellent precision \cite{Shen:2014vra}.

In relativistic hydrodynamics the space-time evolution of the QGP medium is followed by means of the conservation equations
\begin{equation}
  \partial_\mu T^{\mu\nu} = 0
  \label{eq:conservation}
\end{equation}
for the energy-momentum tensor; in the case of a viscous fluid it is given by
\begin{equation}
  T^{\mu\nu} = e \, u^\mu u^\nu  - \Delta^{\mu\nu} (P + \Pi) + \pi^{\mu\nu},
  \label{eq:Tmunu_hydro}
\end{equation}
where $u^\mu$ is the fluid flow velocity, $e$ and $P$ are the energy density and the equilibrium pressure in the local rest frame (LRF) of the fluid, $\Pi$ and $\pi^{\mu\nu}$ are the bulk viscous pressure and the shear stress tensor representing the viscous corrections with respect to the ideal fluid and $\Delta^{\mu\nu}=g^{\mu\nu}-u^\mu u^\nu$ is the projector onto the space orthogonal to $u^\mu$, being $g^{\mu\nu}=\mathrm{diag}(1,-1,-1,-1)$ the metric tensor.
In a general frame the fluid 4-velocity is given by $u^\mu =\gamma\ (1,\boldsymbol{\beta})$, with $\boldsymbol{\beta}=(\beta_x,\beta_y,\beta_z)$ being the 3-velocity of the considered fluid element and $\gamma = 1/\sqrt{1-\boldsymbol{\beta}^2}$ the corresponding Lorentz factor.
In order to solve Eq.~\eqref{eq:conservation} the initial conditions for $u^\mu$, $e$, $P$, $\Pi$ and $\pi^{\mu\nu}$ should be provided. 
We use the method from Ref.~\cite{Liu:2015nwa} for determining $\pi^{\mu\nu}$.
The numerical implementation of viscous hydrodynamics in VISHNew calculates the time evolution of the viscous corrections through the second-order Israel-Stewart equations \cite{Israel:1979wp, ISRAEL1976213} in the 14-momentum approximation, providing a set of relaxation-type equations \cite{Denicol:2014vaa}:
\begin{subequations}
  \label{eq:relaxation}
  \begin{align}
    \tau_\Pi \dot{\Pi} + \Pi &=
      -\zeta \theta - \delta_{\Pi\Pi} \Pi\theta + \phi_1 \Pi^2 \nonumber \\
      &\qquad + \lambda_{\Pi\pi} \pi^{\mu\nu} \sigma_{\mu\nu}
      + \phi_3 \pi^{\mu\nu}\pi_{\mu\nu}, \\
    \tau_\pi \dot{\pi}^{\langle \mu\nu \rangle} + \pi^{\mu\nu} &=
      2\eta\sigma^{\mu\nu} + 2\pi_\alpha^{\langle \mu} \omega^{\nu \rangle \alpha}
      - \delta_{\pi\pi} \pi^{\mu\nu} \theta \nonumber \\
      &\qquad + \phi_7 \pi_\alpha^{\langle \mu} \pi^{\nu \rangle \alpha}
      - \tau_{\pi\pi} \pi_\alpha^{\langle \mu}\sigma^{\nu \rangle \alpha} \nonumber \\
      &\qquad + \lambda_{\pi\Pi} \Pi \sigma^{\mu\nu} + \phi_6 \Pi \pi^{\mu\nu},
  \end{align}
\end{subequations}
where $\omega^{\mu\nu}\equiv(\nabla^{\mu}u^{\nu}-\nabla^{\nu}u^{\mu})/2$ is the vorticity tensor, $\sigma^{\mu\nu}\equiv\nabla^{\left\langle\mu\right.} u^{\left.\nu\right
\rangle}$ the velocity stress tensor and $\theta\equiv\nabla_\mu u^\mu$ the scalar expansion rate, being $\nabla_{\mu}=\Delta_{\mu}^{\nu}\partial_{\nu}$ the projected spatial gradient and having used the notation $A^{\langle\mu\nu\rangle }\equiv \Delta _{\alpha\beta}^{\mu \nu }A^{\alpha \beta }$, with $\Delta_{\alpha \beta }^{\mu \nu }\equiv (\Delta_{\alpha }^{\mu}\Delta _{\beta }^{\nu }+\Delta_{\beta }^{\mu }\Delta_{\alpha}^{\nu}-2/3\Delta^{\mu\nu}\Delta_{\alpha \beta })/2$.
The transport coefficients $\eta$ and $\zeta$ are the shear and bulk viscosities and $\tau_{\pi}$ and $\tau_{\Pi}$ are their associated relaxation times.

The hydrodynamic simulations utilize a parametrization of the $\eta/s(T)$ obtained within PHSD that is very similar to the temperature dependence of $\eta/s$ determined via the Bayesian analysis of the available experimental data \cite{Bernhard:2019bmu}. 
Regarding the bulk viscosity, while the DQPM predicts rather large $\zeta/s$ at $T_C$, the  maximum $\zeta/s$ that the hydrodynamical model can handle is much smaller than that from PHSD simulations. 
In the present VISHNew calculations we used $\zeta/s$ obtained from the Bayesian analysis of experimental data \cite{Bernhard:2019bmu}.
The influence of $\zeta/s$ on the medium evolution in hydrodynamic calculations have been investigated in Ref.~\cite{Xu:2017pna}.

For the other transport coefficients we use analytic results derived for a gas of classical particles in the limit of small but finite masses \cite{Denicol:2014vaa}.
The hydrodynamic equations of motion must be closed by an EoS $P = P(e)$. We use a modern QCD EoS based on continuum extrapolated lattice calculations at zero baryon density from the HotQCD collaboration \cite{HotQCD:2014kol} and then blended into a hadron resonance gas EoS in the temperature region {$165 \le T \le 200$~MeV} using a smooth step interpolation function \cite{Moreland:2018gsh}.
In the PHSD the description of the QGP is done within the DQPM model which is based on the lattice QCD EoS from the BMW group \cite{Borsanyi:2012cr, Borsanyi:2013bia}.
The EoS from the both lattice QCD group are very similar.

The hydrodynamic medium is converted into particles as it cools below the $T_{switch}$ isotherm, sampling particles by means of the standard Cooper-Frye algorithm \cite{Cooper:1974mv}.
It preserves the continuity of energy and momentum at the interface between the hydrodynamic regime and the hadronic phase. The latter is simulated with the Ultrarelativistic Quantum Molecular Dynamics (UrQMD) model \cite{urqmd, Bass:1998ca, Bleicher:1999xi}, which describes microscopically the space-time evolution of hadronic matter until the particles stop interacting.

\subsection{Initial conditions}
\label{sec:IC}

In PHSD the simulation starts with the two colliding nuclei approaching one each other at the velocity determined by the beam energy; the participant and spectator regions are selected dynamically by means of the collisions between nucleons of the initial nuclei.
Then, a hot QGP and the hadronic ``corona'' are dynamically generated and evolve according to the proper transport equations, as explained in Sec.~\ref{sec:phsd}.
Moreover, in PHSD there is no equilibrium assumption on the nature of the hot medium during its whole evolution from initial nuclear overlap to final hadronic freeze-out.

A hydrodynamic simulation, such that encoded in VISHNew, starts from a specified initial condition at the thermalization time $t_0$ of the medium. 
While some initialization models generate parametric initial conditions directly at $t_0$ \cite{Miller:2007ri, Kharzeev:2000ph, Kharzeev:2001gp, Moreland:2014oya}, other modellings consist in full initial-state calculations which explicitly treat the pre-equilibrium dynamics \cite{Schenke:2012wb, vanderSchee:2013pia} and generate outcomes that are matched with the hydrodynamic code at $t_0$.
The latter strategy is adopted in this work, using PHSD for extracting the initial conditions for the hydrodynamic simulation. Indeed, the PHSD model describes the early out-of-equilibrium dynamics of relativistic nuclear collisions and, besides being used to simulate the whole collision evolution, it can be used to generate the initial conditions for a subsequent hydrodynamic description of the medium.
In this way, the macroscopic hydrodynamic evolution can be compared to the microscopic transport dynamics starting from the same initial configuration.

The possibility to continue to follow the fireball evolution by means of the hadronic transport description in PHSD and the hadronic afterburner in the second model, as explained in Secs.~\ref{sec:phsd} and \ref{sec:Duke}, permits also to gain information about final particle production and properties, such as the number of charged particles and their distribution in spherocity within the two approaches.

The method to extract the initial conditions for the hydrodynamic evolution from PHSD is the same used in Ref.~\cite{Xu:2017pna}, where the traces of nonequilibrium dynamics in relativistic heavy-ion collisions have been investigated.
We review it here.
First of all, the energy-momentum tensor $T^{\mu \nu}$ should be evaluated in PHSD, which describes the full 3+1D evolution of the medium. 
In order to do that, the spacetime is divided into cells where the energy-momentum tensor is determined in the computational frame from:
\begin{equation}
    T^{\mu \nu}(x) = \sum_{i} \int_0^\infty \frac{d^3p_i}{(2\pi)^3}\ f_i(E_i)\ \frac{p_i^\mu p_i^\nu}{E_i},
    \label{Tmunu_PHSD}
\end{equation}
where $f_i(E)$ is the distribution function corresponding to the particle $i$, $p_i^\mu $ the 4-momentum and $E_i=p_i^0$ is the energy of the particle $i$.

For p+Pb collisions at LHC energy, that are the focus of this work, the cell size is $\Delta x = \Delta y = 0.3$~fm on the transverse plane and $\Delta z = 0.5 \times t/\gamma_{NN}$ scaled by $\gamma_{NN}$ to account for the expansion of the system along the longitudinal direction, being $\gamma_{NN}=\sqrt{s_{NN}}/2m_N$ the Lorentz contraction factor with $\sqrt{s_{NN}}$ the center-of-mass energy for nucleon pair of the collision and $m_N$ the nucleon mass.
We note that the chosen resolution is higher than that used in Ref.~\cite{Xu:2017pna}, because of the smaller size and faster time evolution in proton-nucleus reaction at LHC energy with respect to the nucleus-nucleus collisions at top RHIC energy studied in \cite{Xu:2017pna}.

The energy-momentum tensor of an ideal fluid is diagonal in the LRF of the fluid, where the flow velocity is $u^\mu = (1,0,0,0)$.
In this case the energy density in the cell can be identified with the $T^{00}$ component and the other three diagonal components define the local pressure in the cell that is the same in all directions.
However, in realistic nuclear collisions the fluid is  viscous and anisotropic, hence the components of the pressure along the various directions are different, especially in the early nonequilibrium stage in which we are interested to extract the initial conditions for starting the hydrodynamic evolution.
In this more general case, in order to obtain the quantities ($e,\boldsymbol{\beta}$) from $T^{\mu \nu}$, we have to express them in the LRF of each cell of our space-time grid and diagonalize the energy-momentum tensor.
The Landau matching condition gives the energy density $e$ and the fluid flow velocity $u^\mu$ as, respectively, the timelike eigenvalue and eigenvector of the energy-momentum tensor:
\begin{equation}
  T^{\mu\nu} u_\nu = e u^\mu.
\end{equation}
For details of the calculation see Ref.~\cite{Xu:2017pna}.
The equilibrium pressure can then be determined from the EoS $P = P(e)$ and the bulk viscous pressure from the difference with the total pressure:
\begin{equation}
  P + \Pi = -\frac{1}{3} \Delta_{\mu\nu} T^{\mu\nu}.
\end{equation}
The shear stress tensor can be obtained by rearranging Eq.~\eqref{eq:Tmunu_hydro}:
\begin{equation}
  \pi^{\mu\nu} = T^{\mu\nu} - e u^\mu u^\nu + (P + \Pi) \Delta^{\mu\nu}.
\end{equation}
These quantities allow to initialize the hydrodynamic equations of motion including the viscous corrections.

The PHSD code is based on the parallel ensemble method \cite{Cassing:2009vt}.
The idea of this method is that each nucleon is presented by \textit{NUM} point-like test particles spreading around the center of nucleon by Monte-Carlo according to the nucleon radius. Each of the \textit{NUM} test particles is redistributed over \textit{NUM} different ensembles, such that each ensemble contains \textit{A+A} or \textit{p+A} nucleons, depending on the system under consideration.
NUM parallel ensembles represent initially one ``event''. 
The collisions between particles is allowed only inside one event, while interactions 
on the mean-field level at a given time are computed averaging over the \textit{NUM} events.
The choice of \textit{NUM} has an impact on the fluctuations in the density of particles in the initial phase, but not to their initial correlations.
However, for this study we use a special initialization where the nucleons (centers of the distributions) are located at the same place in each of the \textit{NUM} parallel ensembles. This allows to keep the initial correlations and the choice of NUM has no effect at the initial time; there would be a very mild effect during time evolution due to collisions and mean-field interaction, but this effect is negligible in \textit{pA} collisions due to its short time evolution and especially in the early phase when we extract initial conditions for hydrodynamics.

In the following sections we investigate the difference for the microscopic and macroscopic descriptions of various quantities as a function of time and for different values of the initialization time for the hydrodynamic evolution $t_0$.
We consider a single PHSD event, which consists of $\textit{NUM}=30$ parallel events, and the hydrodynamic events starting with initial conditions extracted from that PHSD event.

%---------------------------------------------------------------
\section{Medium evolution: hydrodynamics versus PHSD}
\label{sec:med}

In this section we compare the microscopic PHSD evolution with the response of the hydrodynamic long-wavelength evolution to the PHSD initial conditions.
We simulate the evolution of the QGP medium by means of the two different models previously explained: the nonequilibrium microscopic approach PHSD and the (2+1)D hydrodynamic macroscopic model VISHNew. The initial condition for the hydrodynamic simulation are determined from PHSD, as discussed previously, at three different initial times: $t_0=0.2$ fm$/c$, $t_0=0.4$ fm$/c$ and $t_0=0.6$ fm$/c$.
The initial flow and viscous corrections from PHSD are included in the initial conditions for the hydrodynamic simulations.
Even though the two models share the same initial conditions at $t_0$, the subsequent evolution may be very different due to the different underlying dynamics.

\subsection{Space-time evolution of energy density, temperature and velocity} 

\begin{figure*}
\centering
\includegraphics[trim={30 20 60 20},clip,width=1.8\columnwidth]{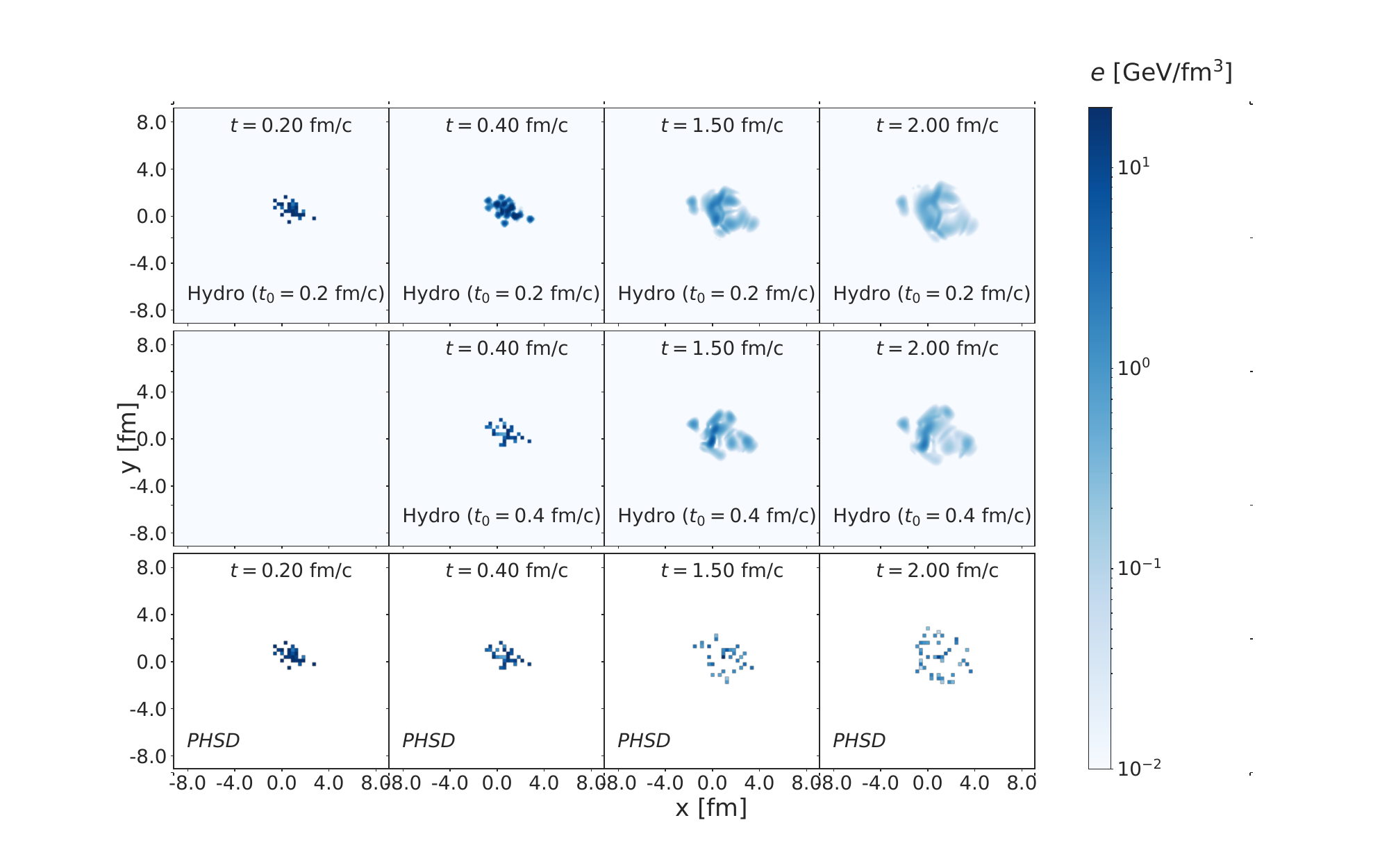}
\\[\baselineskip]% adds vertical line spacing
\includegraphics[trim={30 20 60 20},clip,width=1.8\columnwidth]{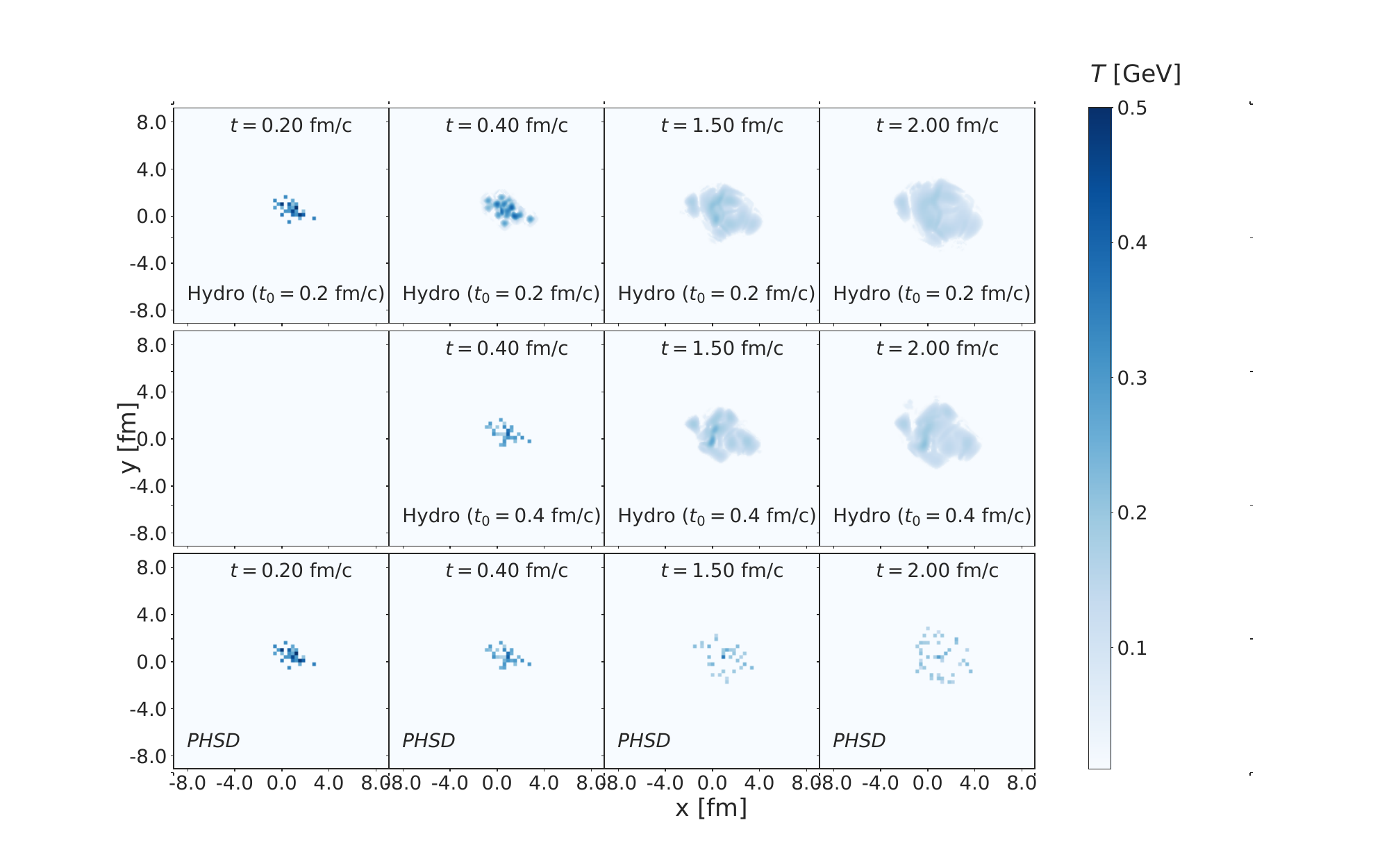}
\caption{Local energy density $e$ (top) and temperature $T$ (bottom) on the transverse plane at $z=0$ of a single event from PHSD (NUM=30) and VISHNew at different times for p+Pb collisions at $\sqrt{s_{NN}}=5.02$ TeV with $b=2$ fm.}
\label{fig:endens_temp}
\end{figure*}

In the top panel of Fig.~\ref{fig:endens_temp} we show the time evolution of the local energy density $e(x,y,z=0)$ in the transverse plane of a single PHSD event (NUM=30) and a single hydrodynamic event at different times for a p+Pb collision at $\sqrt{s_{NN}}=5.02$ TeV with impact parameter $b=2$ fm.
We see that the energy density decreases rapidly as the medium expands. In PHSD the energy density profile is highly inhomogeneous in the transverse plane during the whole time evolution. In the hydrodynamic simulations the energy density profile after $t_0$ becomes smoother than that in PHSD since the hot spots dissolve more efficiently, but still keeps a high degree of inhomogeneity due to the smaller size of the medium produced in p+Pb collisions with respect to heavy-ion reactions; see Figs. 8-9 of Ref.~\cite{Xu:2017pna} for a comparison to the local energy density profiles in Au+Au collisions at $\sqrt{s_{NN}}=200$ GeV.
We notice also a difference of the VISHNew results depending on the initialization time: the energy density in the simulation with $t_0=0.4$ fm$/c$ retains more hot spots than in the simulation stating at 0.2 fm$/c$ because the fireball is already more diluted and the system has less time to level out the spikes coming from the initial conditions before reaching the pseudocritical temperature of the confinement transition.
In the bottom panel we depict the temperature profile obtained by converting the energy density to the temperature given by the lattice QCD equation of state. Similar considerations as done for the energy density hold also for the temperature but with less pronounced variation given the roughly quartic relationship between the two quantities.
\\
We notice also that the energy density and the temperature do not present generally in p+Pb events the gaussian-like shape produced in heavy-ion collisions with the highest values in the centre of the fireball and decreasing values towards the edges. We see from Fig.~\ref{fig:endens_temp} that the hottest spots are randomly distributed depending mainly on the nucleon density encountered by the proton when it hits the lead nucleus.
The differences between the PHSD and the hydrodynamic results can be attributed to a large extent to the fundamentally different dynamics and the way the two models treat the deviations from local equilibrium.

\begin{figure*}
\centering
\includegraphics[trim={30 20 60 20},clip,width=1.8\columnwidth]{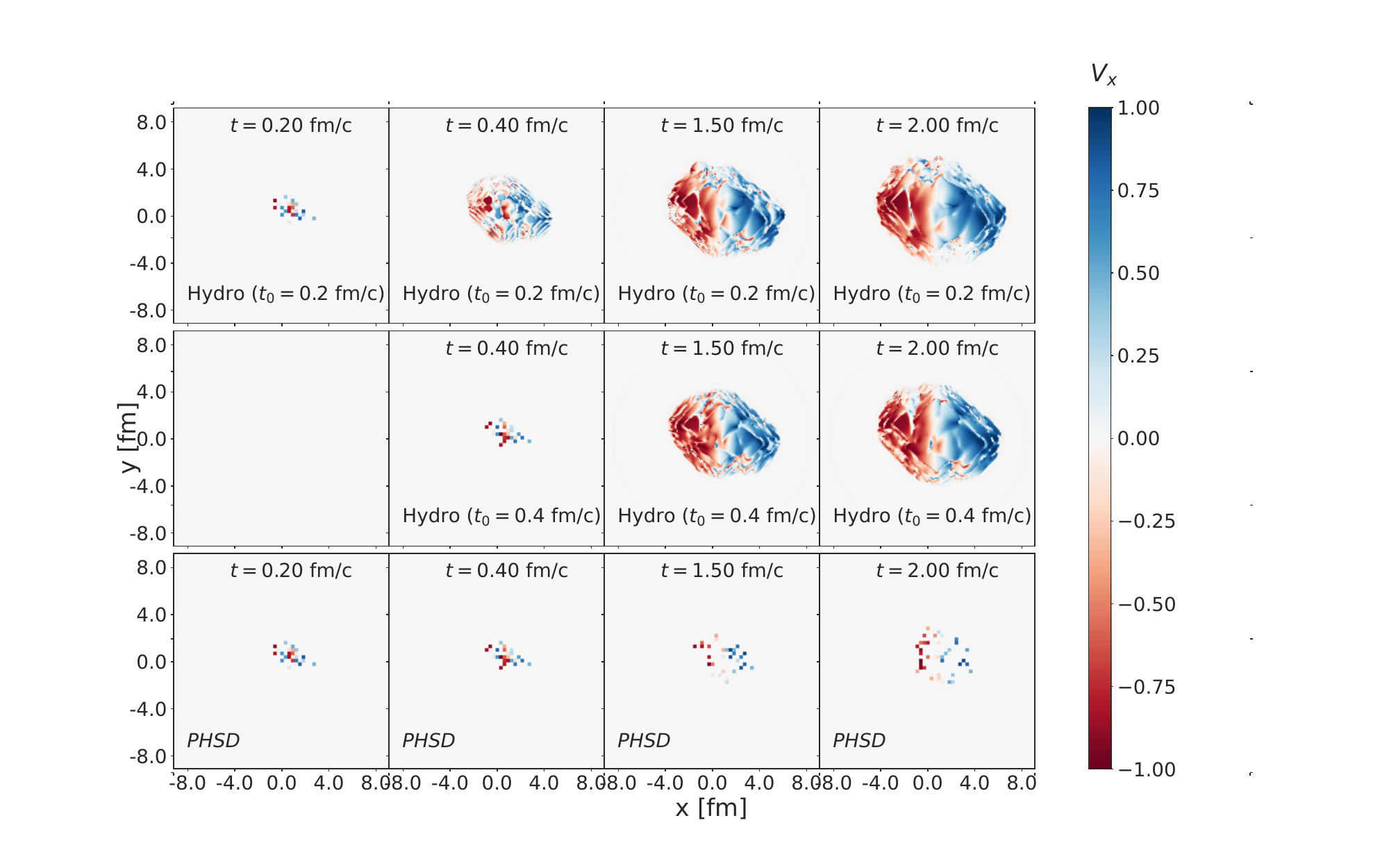}
\\[\baselineskip]% adds vertical line spacing
\includegraphics[trim={30 20 60 20},clip,width=1.8\columnwidth]{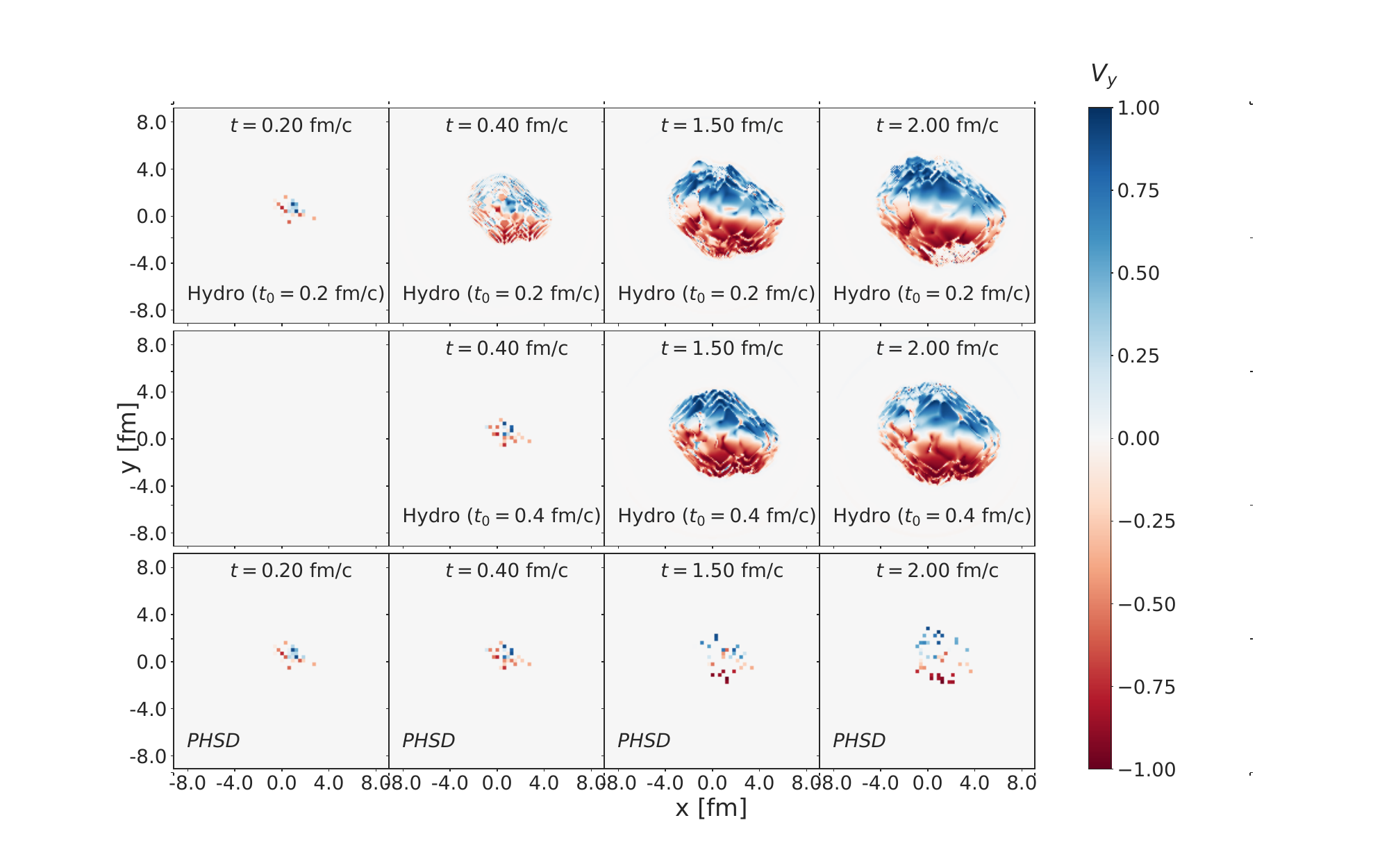}
\caption{Components of the 3-velocity ${\bm \beta}$ along the $x$ direction (top) and $y$ direction (bottom) on the transverse plane at $z=0$ of a single event in PHSD (NUM=30) and VISHNew at different times for p+Pb collisions at $\sqrt{s_{NN}}=5.02$ TeV with $b=2$ fm.}
\label{fig:beta}
\end{figure*}

In Fig.~\ref{fig:beta} we show the time evolution of the transverse components of the velocity $\beta_x$ and $\beta_y$ in the transverse plane for the same event of Fig.~\ref{fig:endens_temp}. The longitudinal velocity $\beta_z$ in the PHSD event is nearly vanishing since we consider a narrow interval in the $z$-direction; in the hydrodynamic event $\beta_z=0$ because we are using a (2+1)D code.
The transverse components of the velocity reach values close to 1 at the edge of the profile, both in VISHNew and in PHSD simulations. In both cases the velocity increases with time, but local fluctuations in a single event are more visible in the PHSD event at later times.

\begin{figure*}
\centering
\includegraphics[trim={30 20 60 20},clip,width=1.8\columnwidth]{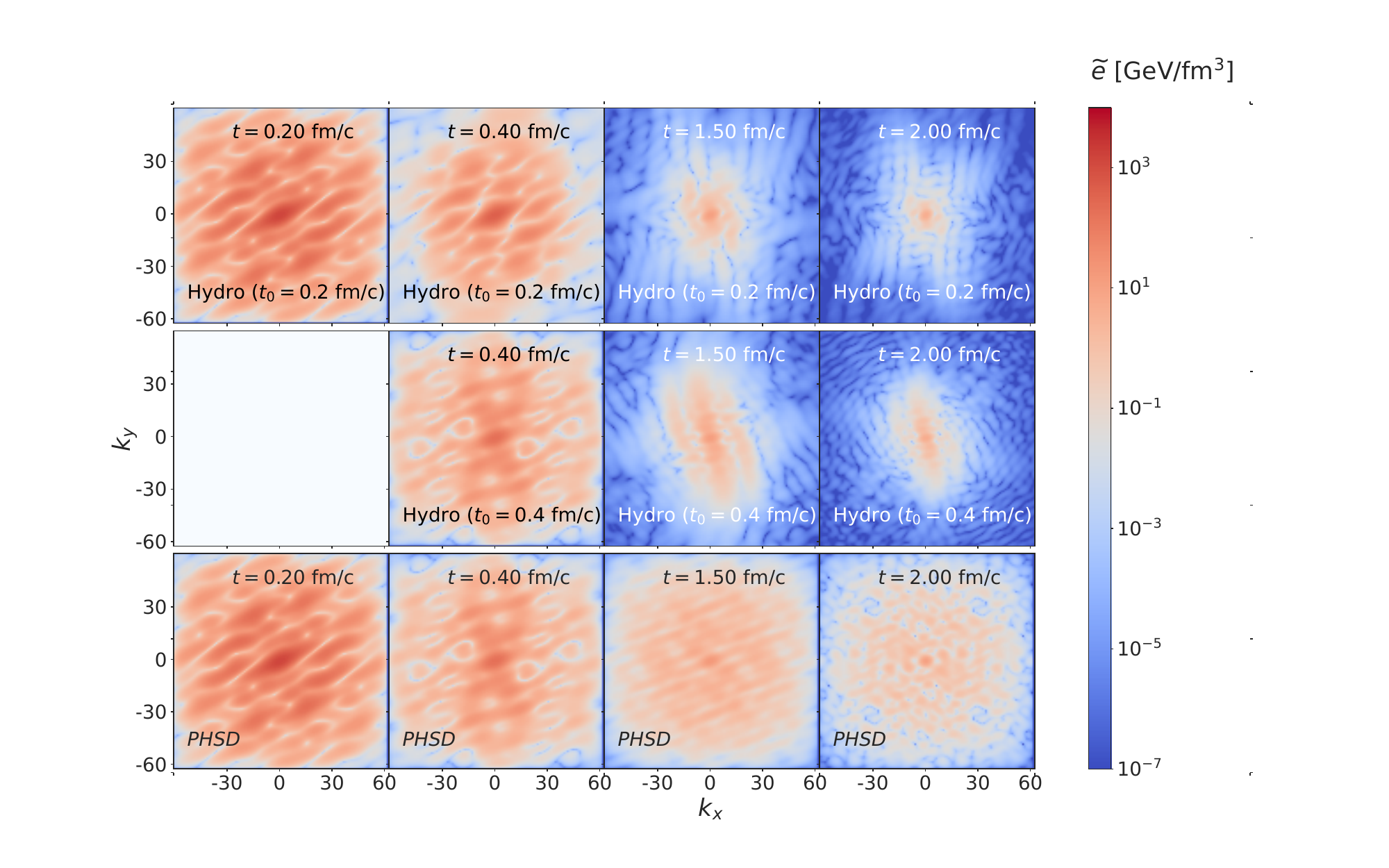}
\caption{Contour plots of the Fourier transform of the energy density $\tilde{e}(x,y,z=0)$ on the transverse momentum plane of a single event from PHSD (NUM=30) and VISHNew at different times for p+Pb collisions at $\sqrt{s_{NN}}=5.02$ TeV with impact parameter $b=2$ fm, obtained from Fig.~\ref{fig:endens_temp}.}
\label{fig:endens_temp_fft}
\end{figure*}

\begin{figure*}
\centering
\includegraphics[trim={0 10 50 0},clip,width=2\columnwidth]{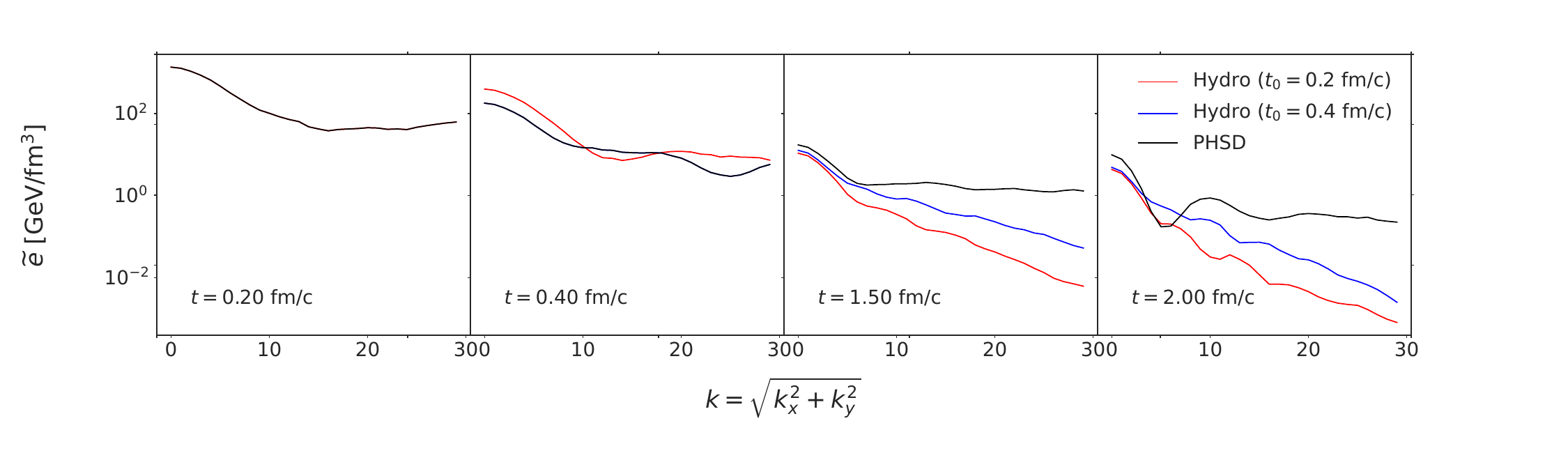}
\caption{Radial distribution of the Fourier modes of the energy density of a single event from PHSD (NUM=30) and VISHNew at different times for p+Pb collisions at $\sqrt{s_{NN}}=5.02$ TeV with impact parameter $b=2$ fm.}
\label{fig:endens_temp_fft_radial}
\end{figure*}

In order to better quantify the inhomogeneity of the medium we compute the Fourier transform of the energy density shown in the top panel of Fig.~\ref{fig:endens_temp}. For a discrete spatial grid with an energy distribution as $e(x,y)_{m\times n}$ and grid size $(L_x, L_y)=(0.1 \,\mathrm{fm}, 0.1 \,\mathrm{fm})$, the Fourier coefficients are given by
\begin{equation}
\tilde{e}(k_x, k_y) =  \frac{1}{m} \frac{1}{n} \sum\limits_{x=0}^{m-1} \sum\limits_{y=0}^{n-1} e(x, y) e^{2\pi i (\frac{x k_x}{L_x m} + \frac{y k_y}{L_y n})}\, .
\end{equation} 
The zero mode $\tilde{e}(k_x=0,k_y=0)$ corresponds to the sum of the energy density values over all grid points, while the higher order coefficients give information about the correlations of the local energy density on different length scales.

The Fourier image $\tilde{e}(k_x,k_y)$ of the energy density is depicted in Fig.~\ref{fig:endens_temp_fft} for both PHSD and hydrodynamics at different times.
In the hydrodynamic evolution, after about 1 fm$/c$, the lower Fourier modes dominate while higher-order coefficients are suppressed: only the global shape of the event survives and shorter wavelength irregularities are washed out, as expected for a medium with large wavelength structures.
In the PHSD event the short wavelength inhomogeneities are present during all evolution, but with smaller strength with increasing time due to the system dilution. The excitation of higher Fourier modes is inherent to a microscopic nonequilibrium dynamics, such that in PHSD.

The difference between the two approaches in the treatment of short wavelength irregularities can be more easily appreciated from Fig.~\ref{fig:endens_temp_fft_radial}, where the distribution of the Fourier coefficients $\langle\tilde{e}\left(\sqrt{k_x^2+k_y^2}\right)\rangle$ is shown at different times.
We see again that the strength of the Fourier modes is similar in PHSD and VISHnew at the initial times but after the first fm$/c$ the values of shorter wavelength modes rapidly decrease with respect to the zero mode in the hydrodynamic medium, while a high degree of inhomogeneity is maintained in the microscopically evolving medium. 

\subsection{Space-time evolution of the viscous corrections}

\begin{figure*}[!hbt]
\centering
\includegraphics[trim={30 20 50 20},clip,width=1.8\columnwidth]{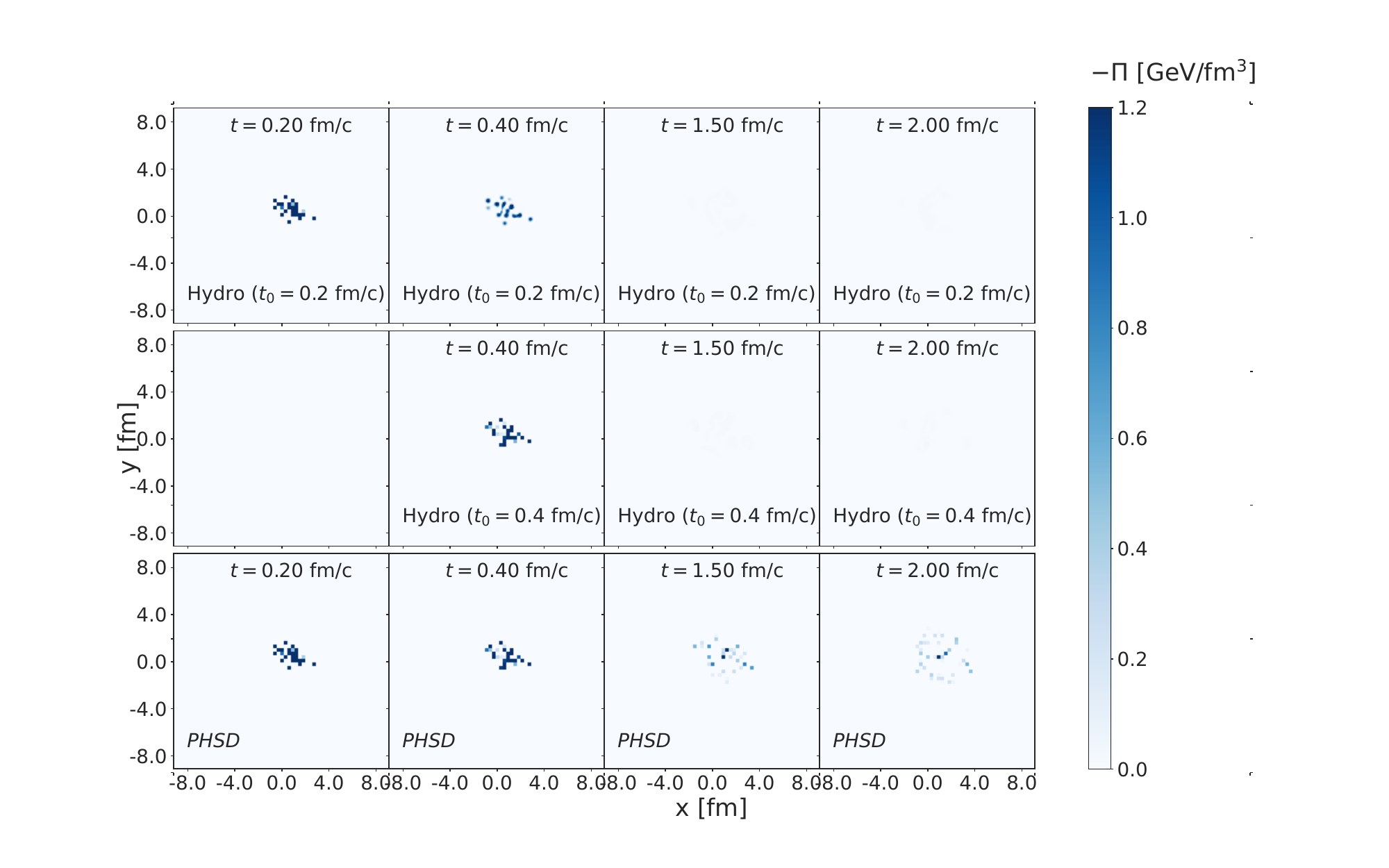}
\includegraphics[trim={30 20 50 20},clip,width=1.8\columnwidth]{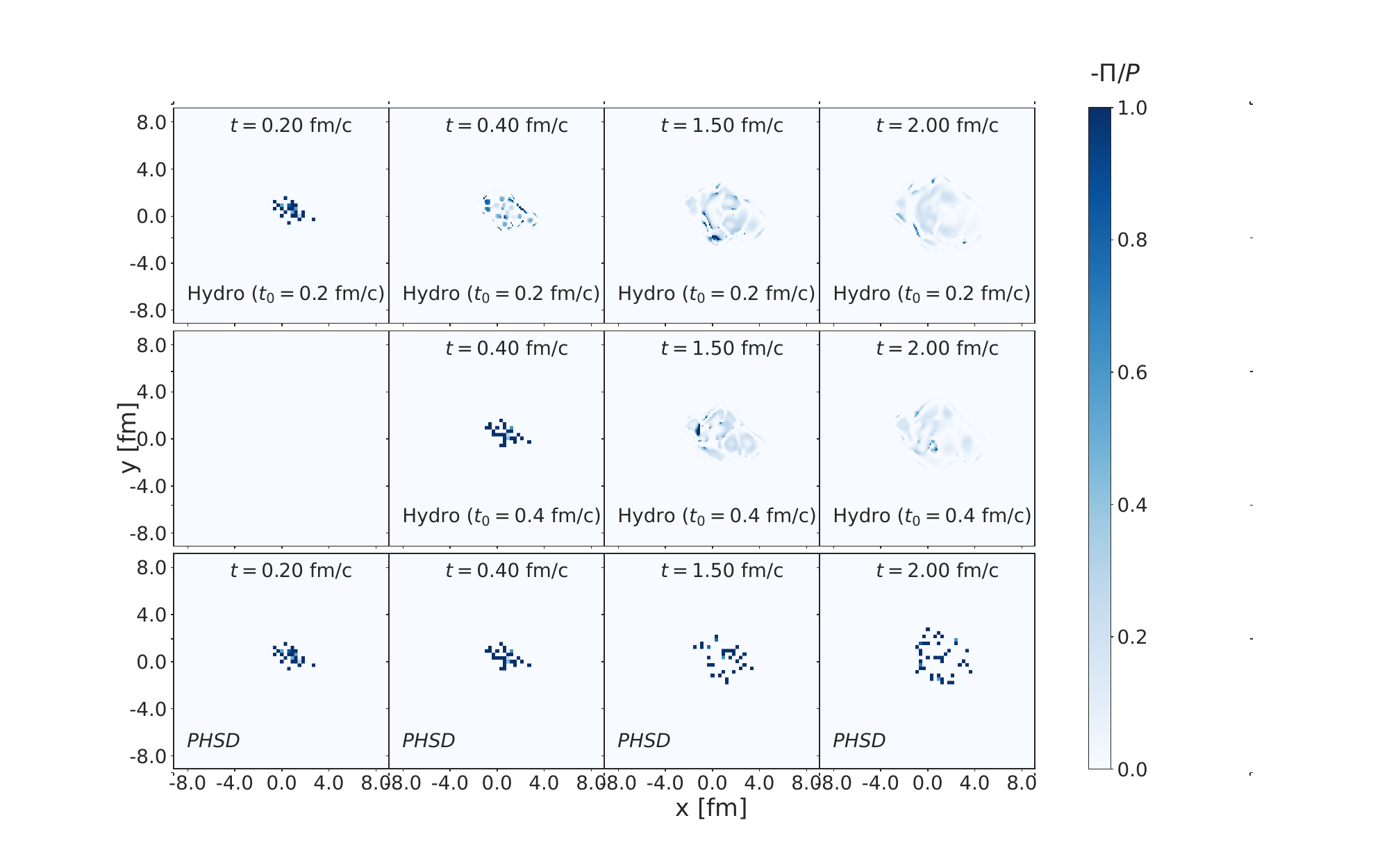}
\caption{Bulk viscous pressure $\Pi$ (top) and ratio of $\Pi$ over equilibrium pressure $P$ (bottom) on the transverse plane at $z=0$ of a single event from PHSD (NUM=30) and VISHNew at different times for p+Pb collisions at $\sqrt{s_{NN}}=5.02$ TeV with $b=2$ fm.}
\label{fig:bvp}
\end{figure*}

In this section we focus on the viscous corrections.
The bulk viscous pressure $\Pi$ is given by
\begin{equation}
\Pi=-\dfrac{1}{3}\mathrm{Tr}\left(\Delta_{\mu\nu}T^{\mu\nu}\right)-P.
\end{equation}

In the top panel of Fig.~\ref{fig:bvp} we show the time evolution of the bulk viscous pressure $-\Pi(x,y,z=0)$ in the transverse plane of a single PHSD event (NUM=30) and a single hydrodynamic event at different times for a p+Pb collision at $\sqrt{s_{NN}}=5.02$ TeV with impact parameter $b=2$ fm.
We see that in VISHNew the bulk viscous pressure approaches zero far more quickly than in the PHSD evolution.
A similar behaviour holds for the components of the shear stress tensor $\pi_{11}$, $\pi_{22}$ and $\pi_{12}$.
In the bottom panel we present the ratio of $-\Pi$ over the equilibrium pressure $P$. We see that the ratio is close to 1 in a single PHSD event during all the evolution, meaning that the deviations from thermal equilibrium are very large; in fact, $-\Pi/P\gtrsim 0.8$ in most of the droplets at $t=2$ fm$/c$. In the hydrodynamic simulation the value of $-\Pi/P$ is close to 1 at the starting time due to the initial condition extracted from PHSD but then it experiences a quick decrease becoming less than 0.3 after about 1 fm$/c$. Hence, the large deviations from equilibrium present both in PHSD and VISHNew at the initial time of a proton-nucleus collision persist during all the transport evolution while they are strongly reduced in the hydrodynamic case, even though some spots where the bulk contribution to pressure is of the same order as the equilibrium pressure still appear at later times also in the latter approach.

\begin{figure}[!hbt]
\centering
\includegraphics[trim={0 0 35 35},clip,width=\columnwidth]{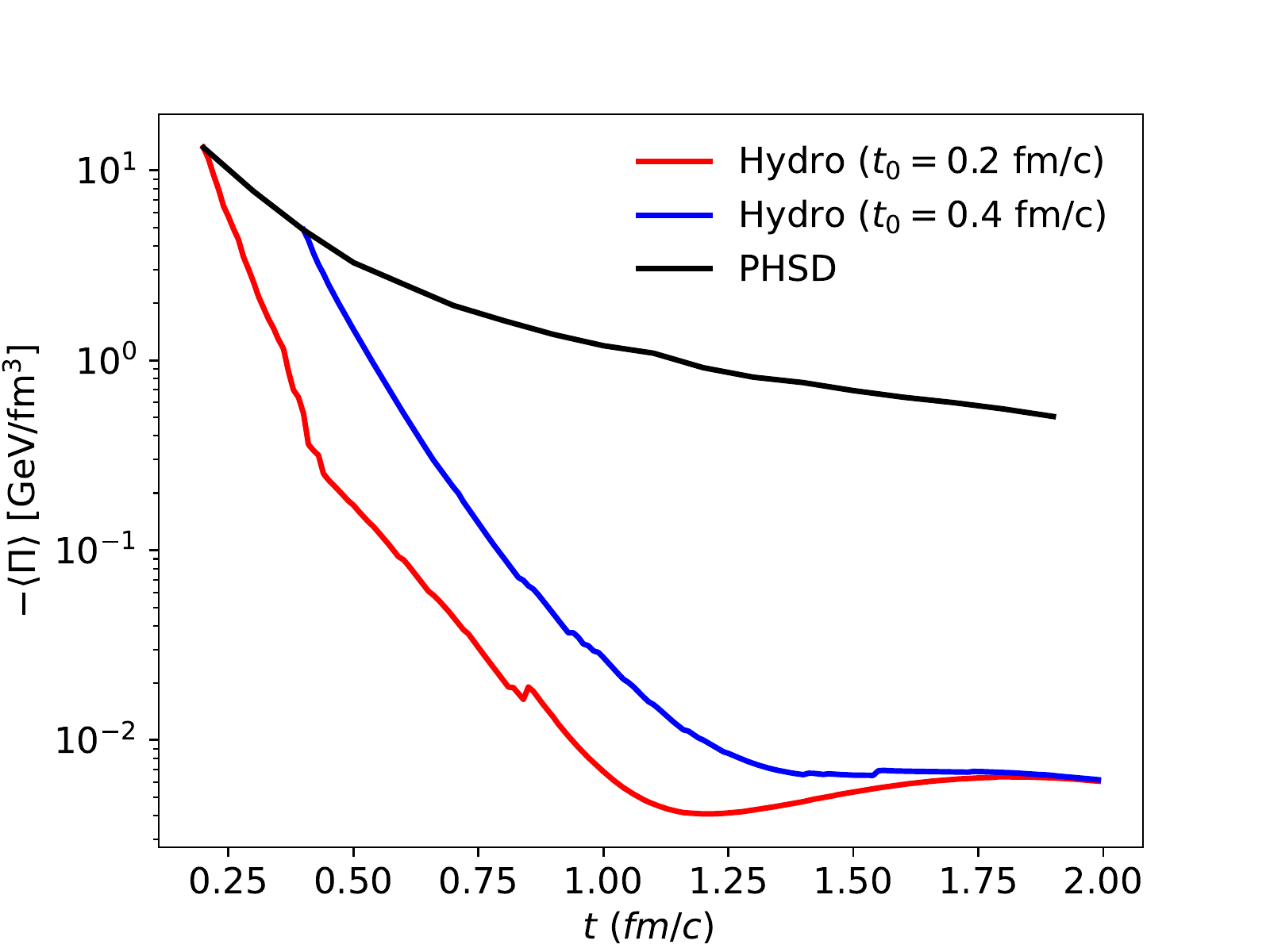}
\caption{Time evolution of the bulk viscous pressure -$\Pi$ averaged over the transverse plane at $z=0$ and weighted by the energy density of 100 PHSD events (black line) and 100 hydrodynamic events (red and blue lines) for p+Pb collisions at $\sqrt{s_{NN}}=5.02$ TeV with $b=2$ fm.}
\label{fig:Pai_t}
\end{figure}

In Fig.~\ref{fig:Pai_t} we depict the absolute value of the event-average of $\langle\Pi\rangle$, defined as the bulk viscous pressure $\Pi$ averaged over the transverse plane weighted with the energy density:
\begin{equation}
    \langle\Pi\rangle=\frac{\int d^2\x_T\,\Pi\,e(x,y)}{\int d^2\x_T\,e(x,y)}.
\end{equation}
The black line corresponds to the PHSD result whereas the red and blue curves are obtained with hydrodynamic simulations starting from PHSD initial conditions at different times.
$\Pi$ is initially very large and negative due to the large initial expansion rate. The magnitude of the bulk viscous pressure in PHSD experiences a power-law decay and at $t=2$ fm$/c$, corresponding roughly to the end of the QGP evolution in the hydrodynamic simulations, has still a value of about 0.5 GeV/fm$^3$. In the hydrodynamic case $\Pi$, which has initially the same value of PHSD at $t_0$ of the simulation, drops very fast with respect to PHSD and approaches quickly zero within about half fm$/c$.
The dissimilarity between the results within the two approaches may be partially due to the differences in the temperature dependence of the bulk viscosity, which mainly drives the evolution of $\Pi$ in hydrodynamics. Indeed, in PHSD $\zeta/s$ is larger than in hydrodynamics \cite{Xu:2017pna}, since the latter can not deal with the large bulk viscosity.
In both cases the evolution of the bulk viscous pressure in small systems is quicker than that in heavy-ion collisions (see, e.g., Ref.~\cite{Song:2009rh}).
We notice that the two lines from hydrodynamic simulations started at different times are very different at the beginning (being them equal to the value in PHSD at that time) but then relax to the same trajectory within 2 fm$/c$. This memory loss of the initial conditions resembles the attractor behaviour \cite{Heller:2015dha}; a similar behaviour is seen in heavy-ion collisions after 1-2 fm$/c$, as shown in Ref.~\cite{Song:2009rh}.

\subsection{Time evolution of the spatial and momentum anisotropy}

In this section we study the medium response to the initial spatial anisotropies.
Within the hydrodynamic framework the initial spatial gradients are transformed via hydrodynamic pressure into momentum anisotropies that can be quantified by means of the Fourier coefficients of the azimuthal particle distribution.
In particular, in the hydrodynamic description of heavy-ion collisions the spatial anisotropies lead to substantial elliptic flow, that is the second Fourier harmonics.
In experiments only the final state particle spectra are accessible, while we can study the space-time evolution of the spatial and momentum anisotropy of the medium.

The spatial anisotropy of the matter distribution is quantified by the eccentricity coefficients $\epsilon_n$ defined as
\begin{equation}
\epsilon_n \exp({i n \Phi_n}) = - \frac{\int r\,dr\, d\phi\, r^n \exp({in\phi})\, e(r, \phi)}{\int r\,dr\, d\phi\, r^n\, e(r, \phi)}
\end{equation}
where $e(r, \phi)$ is the local energy density in the transverse plane. 
\\
The second-order coefficient $\epsilon_x\equiv\epsilon_2$ is also called ellipticity and is to leading order the origin of the elliptic flow $v_2$. It can be simplified to
\begin{equation}
\epsilon_x = \frac{\sqrt{\{r^2 \cos(2\phi)\}^2 + \{r^2 \sin(n\phi)\}^2}}{\{r^2\}}
\end{equation}
where $\{\ldots\} = \int d^2\x_T (\ldots) e(x,y)$ describes an event-averaged quantity weighted by the local energy density $e(x,y)$ \cite{Qiu:2011iv}.

The evolution of the azimuthal anisotropy in momentum space can be followed computing the integrated transverse stress tensor $[T^{ij}]_s = ([T^{xx}]_s, [T^{xy}]_s, [T^{yy}]_s)$, where 
\begin{equation}
\left[\ldots\right]_s \equiv \int d^2\x_T u^\tau (\ldots)
\end{equation}
denotes an integral over the transverse plane without the energy weight, as $T^{\mu\nu}$ is already ``energy weighted''.
We use the following definition of momentum ellipticity \cite{Kurkela:2018vqr}
\begin{equation}
\epsilon_p=\frac{\sqrt{( \left[T^{xx}\right]_s-\left[T^{yy}\right]_s)^2+4\left[T^{xy}\right]_s^2}}{\left[T^{xx}\right]_s+\left[T^{yy}\right]_s} \, ,
\label{eq:epspprime}
\end{equation}
which provides a measure of the elliptic flow as a function of time.
Here the energy-momentum tensor includes the viscous corrections from  $\pi^{\mu\nu}$ and $\Pi$.

\begin{figure}
\centering
\begin{overpic}[abs,unit=1mm,width=\columnwidth,trim={10 0 30 35},clip]{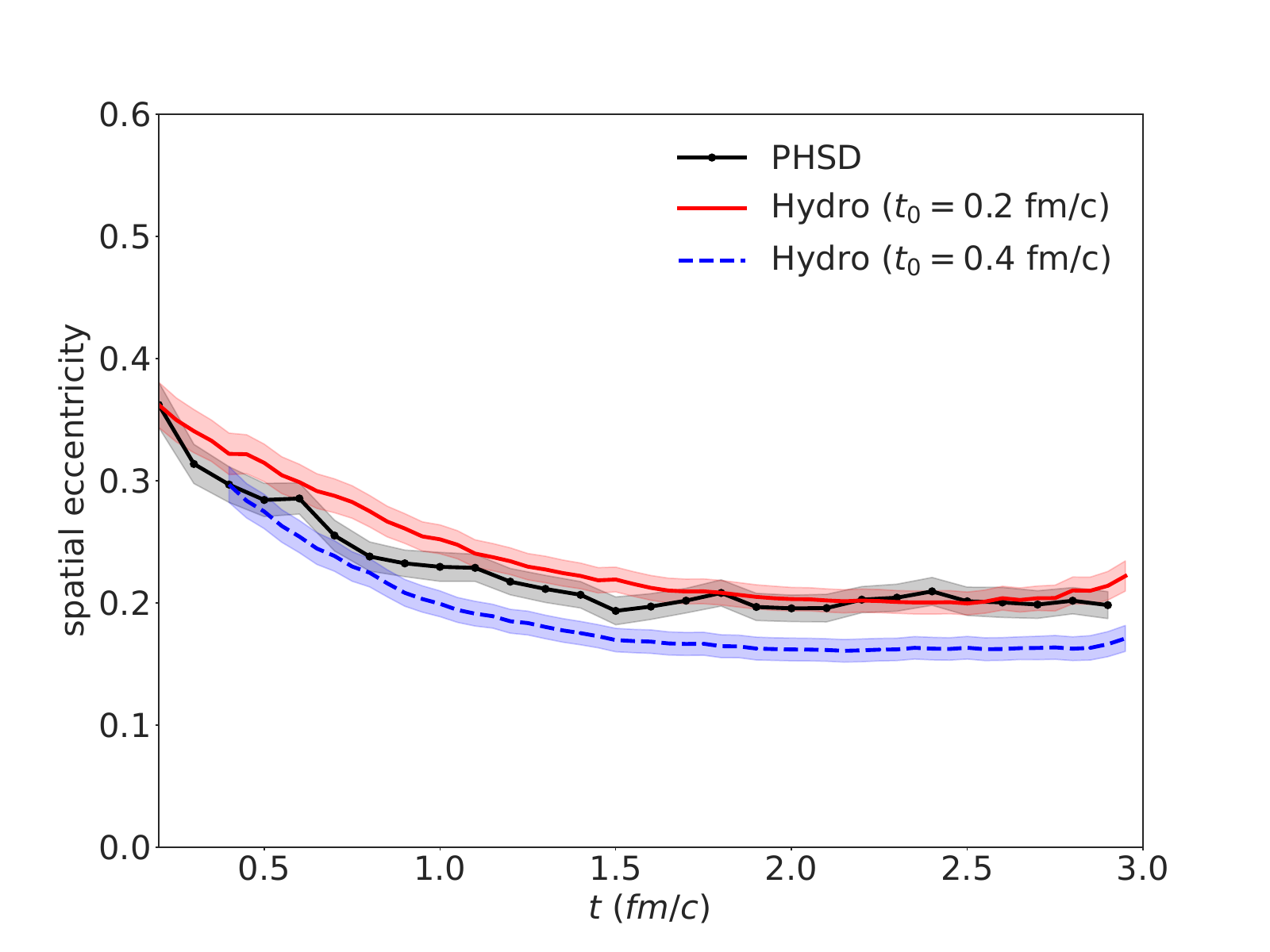}
\put(15,55){\textbf{(a)}}
\end{overpic}
%\par\bigskip % force a bit of vertical whitespace
\begin{overpic}[abs,unit=1mm,width=\columnwidth,trim={10 0 30 35},clip]{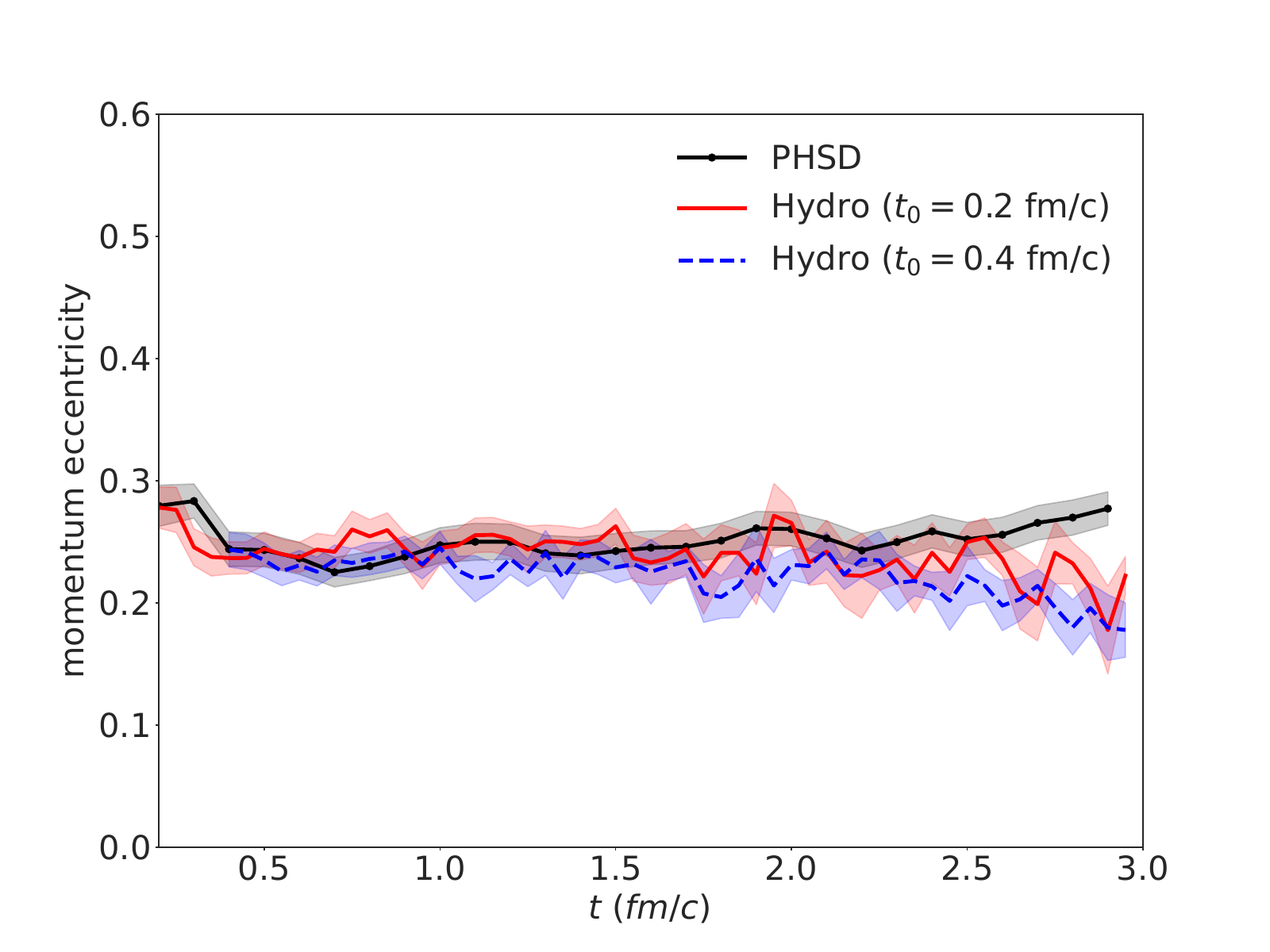}
\put(15,55){\textbf{(b)}}
\end{overpic}
\caption{Temporal evolution of the event-averaged spatial eccentricity $\epsilon_x$ (a) and momentum eccentricity $\epsilon_p$ (b) of 100 events from PHSD and VISHNew for p+Pb collision at $\sqrt{s_{NN}}=5.02$ TeV with impact parameter $b=2$ fm.}
\label{fig:Ecc_b2}
\end{figure}

The PHSD model naturally produces initial state fluctuations due to its microscopic dynamics. We therefore apply event-by-event hydrodynamics and the eccentricities are averaged over many events.
In Fig.~\ref{fig:Ecc_b2} we show the time evolution of the event-by-event averaged ellipticity $\langle\epsilon_x\rangle$ (a)   and momentum eccentricity $\langle\epsilon_p\rangle$ (b) in p+Pb collision at $\sqrt{s_{NN}}=5.02$ TeV with impact parameter $b=2$ fm for both PHSD and VISHNew approaches.
For computing the $\epsilon_x$ the fireball has been centered at $t=0.2$ fm$/c$.
For the hydrodynamic medium description we consider simulations with pre-equilibrium flow in the initial conditions which leads to a finite momentum anisotropy at $t_0$. 
The shaded bands are obtained by calculating the standard error.
We notice that the spatial eccentricity $\epsilon_x$ decreases with time saturating at a value of about 0.2.
For the momentum eccentricity $\epsilon_p$ we find that it is almost constant during the whole time evolution.
This is very different with what has been seen in non-central heavy-ion collisions \cite{Xu:2017pna}, where the momentum eccentricity clearly increases with time as the pressure transforms the spatial anisotropy in collective flow.

\section{Event topology: hydrodynamics versus PHSD}

When moving from the study of heavy-ion collisions to that of small colliding systems, it becomes increasingly important to classify events according not only to multiplicity but also to other observables, in order to have a multi-differential categorization of events.
Transverse spherocity is an observable capable to separate the events based on their geometrical shapes. Recent experimental results from ALICE Collaboration highlight the utility of transverse spherocity as event classifier in small systems for studies related to phenomena, such as the development of collective flows and the enhanced production of strange hadrons \cite{Nassirpour:2020owz}, which are considered signatures of the formation of the quark-gluon plasma and were previously attributed only to heavy-ion collisions.
In this section we study the event topology in the hydrodynamic and PHSD description of the system produced in ultra-relativistic proton-nucleus collisions.

While in the previous section we focused on collisions at fixed impact parameter in order to study the medium evolution in a single event from the microscopic and macroscopic perspectives, in this section we show results obtained simulating the PHSD events without restriction on the impact parameter but with random collisions of the proton towards the lead nucleus according to the geometrical probability. From this minimum bias event sample\footnote{Here, the minimum bias events are those events where no selection on charged-particle multiplicity and/or spherocity is applied.} initial conditions at three different times are extracted for starting the hydrodynamic evolution with the VISHNew code. Then, when the switching temperature is reached in the hydrodynamic evolution (as explained in Sec.~\ref{sec:Duke}) the Cooper-Frye procedure allows to describe hadronization and move into the hadronic stage. From these minimum bias events a centrality selection is done within each of the two approaches according to charged-particle multiplicity in the pseudorapidity range $|\eta|<0.5$ and the 5\% most central events are extracted\footnote{In PHSD it is possible to select centrality from the charged particles in the pseudorapidity range corresponding to the V0A detector of ALICE experiment, as usually done by the ALICE Collaboration for obtaining the experimental data in the various centrality classes. However, this is not possible within the (2+1)D hydrodynamic code VISHNew, hence we restrict to the region $|\eta|<0.5$ in both PHSD and hydrodynamics for determining the 0-5\% centrality class considered in this work.}.\\
All results shown in this section are obtained considering charged particles at midrapidity $|\eta|<0.5$.  The determination of the centrality classes in far-from-midrapidity regions for a more direct comparison with the experimental data is left for a future study.

%---------------------------------------------------------------
\subsection{Charged particle distributions}

First we analyse the distributions of charged particles in minimum bias and 5\% central events within PHSD and the hydrodynamic simulations.

\begin{figure}
\centering
\begin{overpic}[abs,unit=1mm,width=0.95\columnwidth,trim={0 0 35 20},clip]{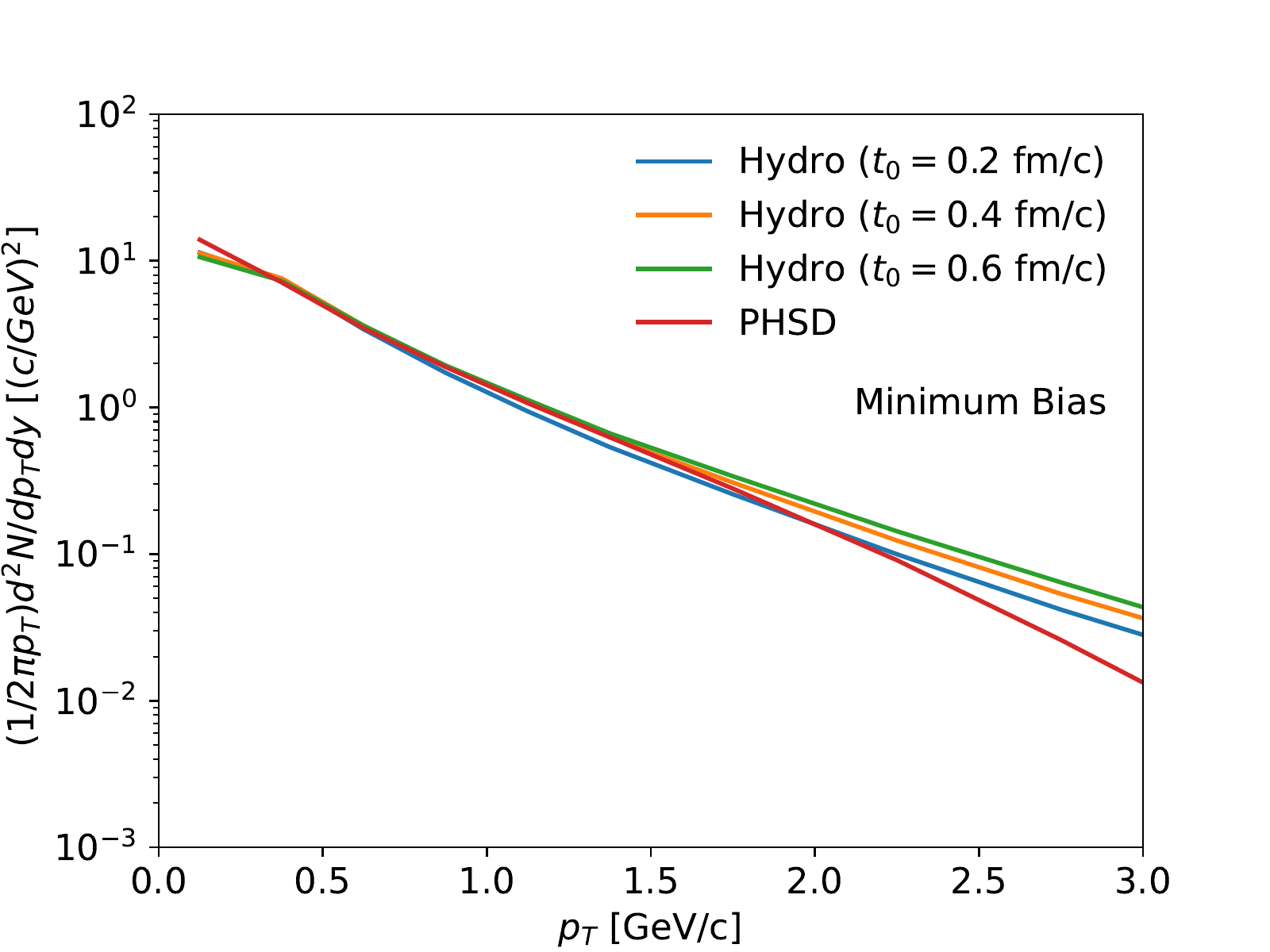}
\put(15,15){\textbf{(a)}}
\end{overpic}
%\par\bigskip % force a bit of vertical whitespace
\begin{overpic}[abs,unit=1mm,width=0.95\columnwidth,trim={0 0 35 20},clip]{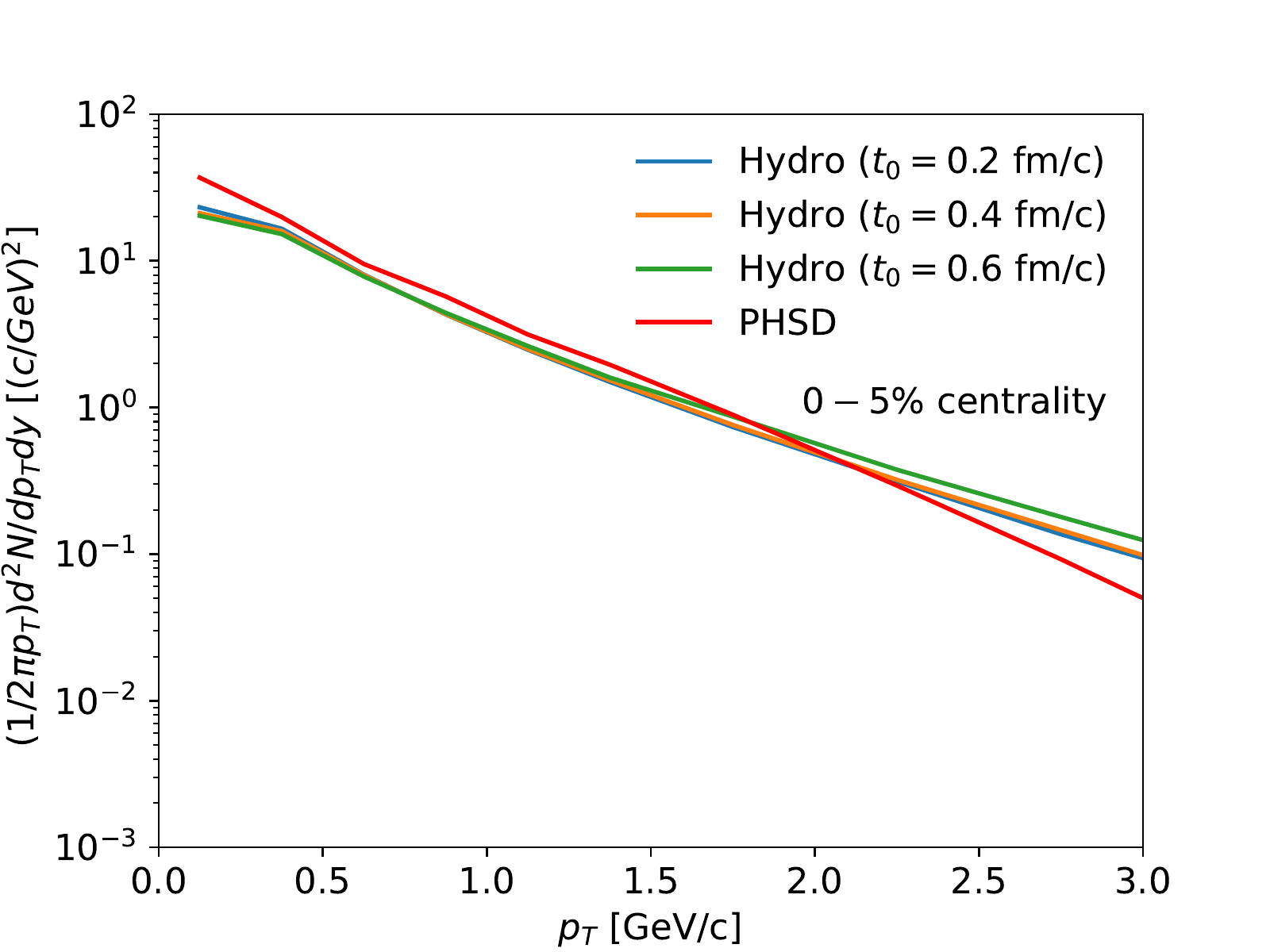}
\put(15,15){\textbf{(b)}}
\end{overpic}
\caption{Transverse momentum spectra of charged particles at midrapidity ($|\eta|<0.5$) for PHSD events (red line) and hydrodynamics starting from PHSD initial conditions at three different times (green, orange and blue lines). Panel (a) corresponds to minimum bias events, while panel (b) shows 5\% central events (with hydrodynamics starting from minimum bias PHSD events and then the 5\% most central events are selected).}
\label{fig:dNch_dpt}
\end{figure}

In Fig.~\ref{fig:dNch_dpt} we show the transverse momentum spectra of charged particles at midrapidity within the two approaches for minimum bias and high-multiplicity events.
In both cases, the spectrum in the hydrodynamic simulations shows a mild dependence on $t_0$: later starting times correspond to harder spectra, meaning that the system has less time to smoothen the hot spots, as visible in Fig.~\ref{fig:endens_temp}. The presence of hot spots is connected to an enhanced particle production at higher $p_T$ values, as also pointed out in Ref.~\cite{Schenke:2012wb}.
The PHSD results agree fairly well with hydrodynamics for transverse momenta $0.5<p_T<2$ GeV$/c$. At higher $p_T$ the PHSD spectrum is softer than the hydrodynamic one. There is also difference at low $p_T$; part of this region is not considered in our results since we apply a cut $p_T<0.15$ GeV/c as is done by the experiments for achieving a good resolution of the spherocity measurement.

\begin{figure}
\centering
\begin{overpic}[abs,unit=1mm,width=\columnwidth,trim={0 0 35 20},clip]{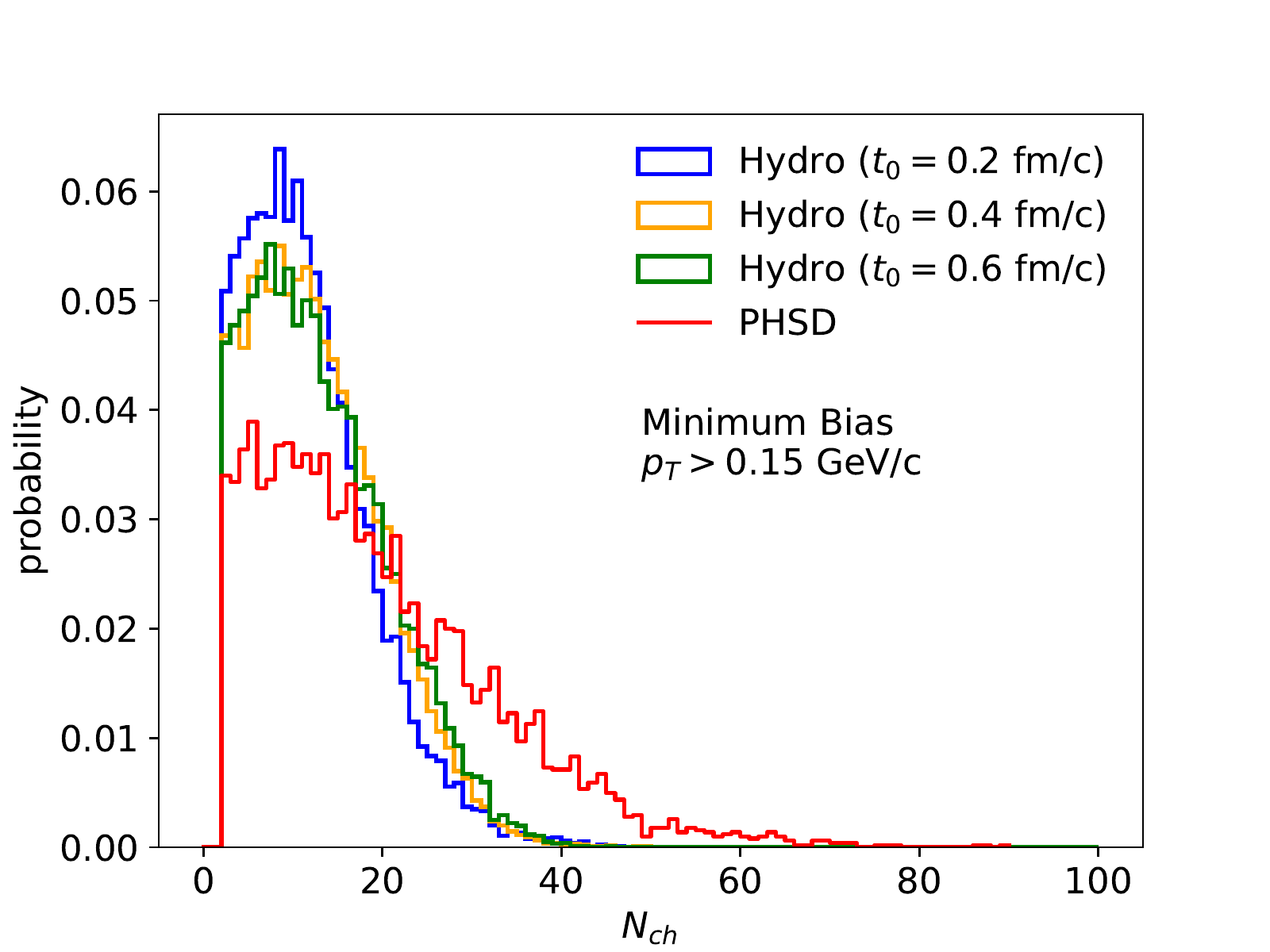}
\put(75,15){\textbf{(a)}}
\end{overpic}
%\par\bigskip % force a bit of vertical whitespace
\begin{overpic}[abs,unit=1mm,width=\columnwidth,trim={0 0 35 20},clip]{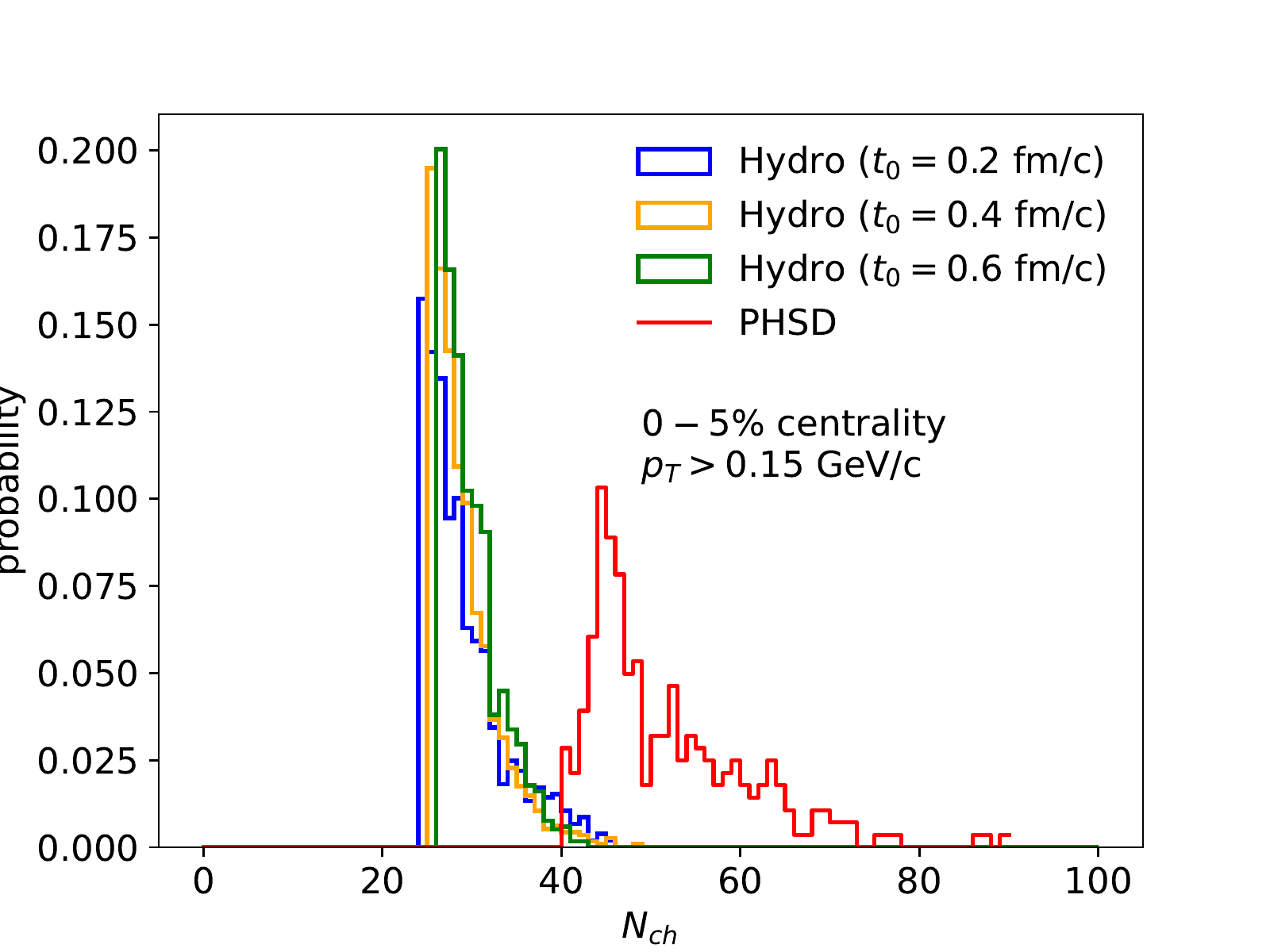}
\put(75,15){\textbf{(b)}}
\end{overpic}
\caption{Event distribution as a function of charged particles for PHSD events (red line) and hydrodynamics starting from PHSD initial conditions at three different times (green, orange and blue lines).
Panel (a) corresponds to minimum bias events, while panel (b) shows 5\% central events (with hydrodynamic starting from minimum bias PHSD events and then the 5\% most central events are selected).}
\label{fig:nch_hydro_phsd}
\end{figure}

In Fig.~\ref{fig:nch_hydro_phsd} we depict the event probability (normalized to 1) as a function of the number of charged particle $N_{ch}$ at midrapidity ($\vert\eta\vert<0.5$) for minimum bias (a) and the 5\% most central events (b). The red lines correspond to PHSD results while the blue, yellow and green curves are obtained with hydrodynamic simulations starting at three different times, $t_0=0.2,0.4,0.6$ fm$/c$ respectively. In the minimum bias case we see that the PHSD distribution has a longer tail towards higher values of $N_{ch}$ while in hydrodynamics more events are produced with a lower multiplicity. By comparing the hydrodynamic results with different starting times we notice that for later $t_0$ the charged-particle distribution moves a little bit closer to the PHSD one but for $t_0=0.4$ fm$/c$ and $t_0=0.6$ fm$/c$ the two lines are quite similar. In the case of the 0-5\% centrality class we see that the difference between PHSD and hydrodynamics increases in terms of event distributions as a function of charged particles. This is expected because we are looking to the tail region at higher multiplicity from panel (a).  

\begin{figure}
\centering
\begin{overpic}[abs,unit=1mm,width=\columnwidth,trim={20 0 35 20},clip]{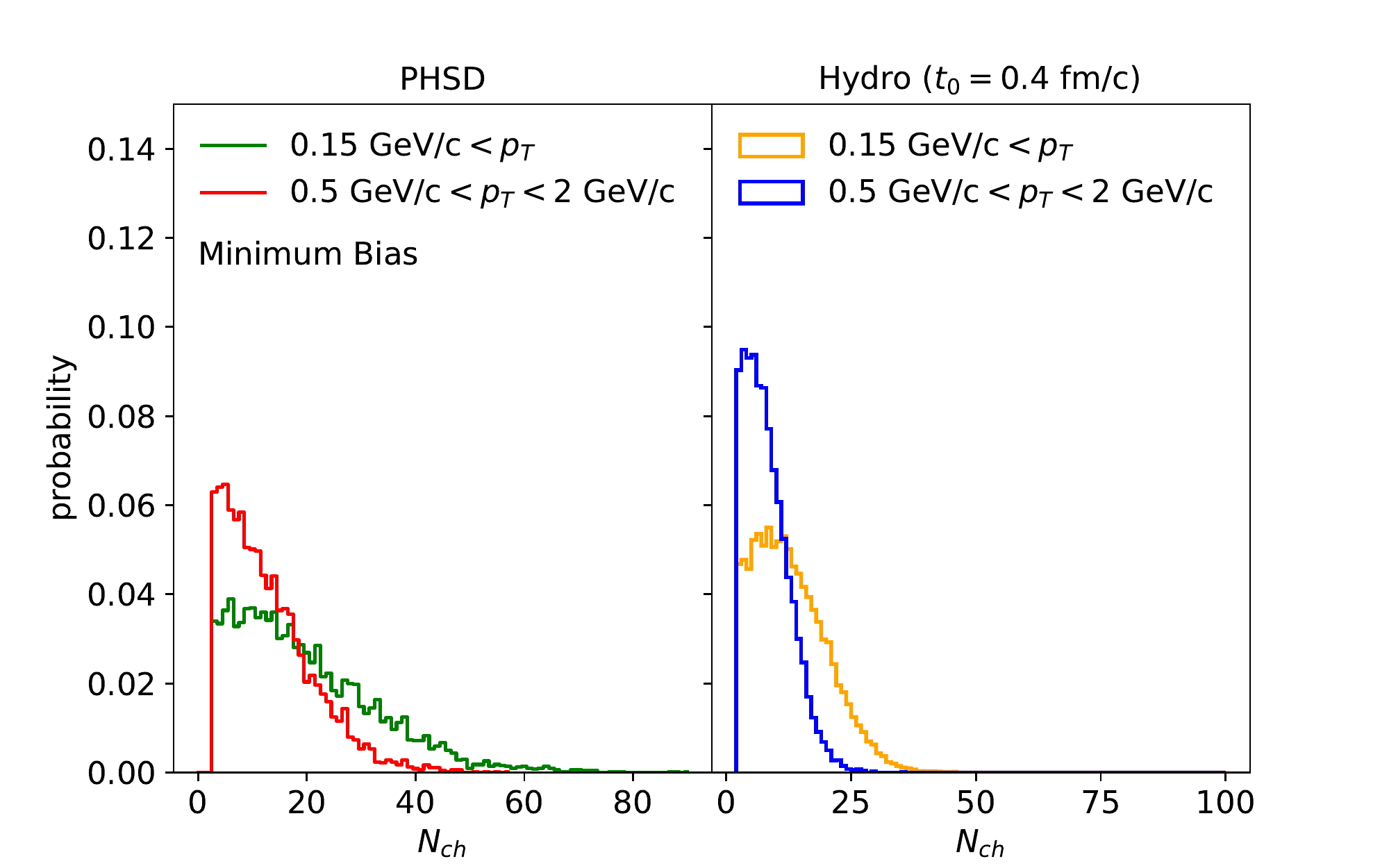}
\put(75,10){\textbf{(a)}}
\end{overpic}
%\par\bigskip % force a bit of vertical whitespace
\begin{overpic}[abs,unit=1mm,width=\columnwidth,trim={20 0 35 20},clip]{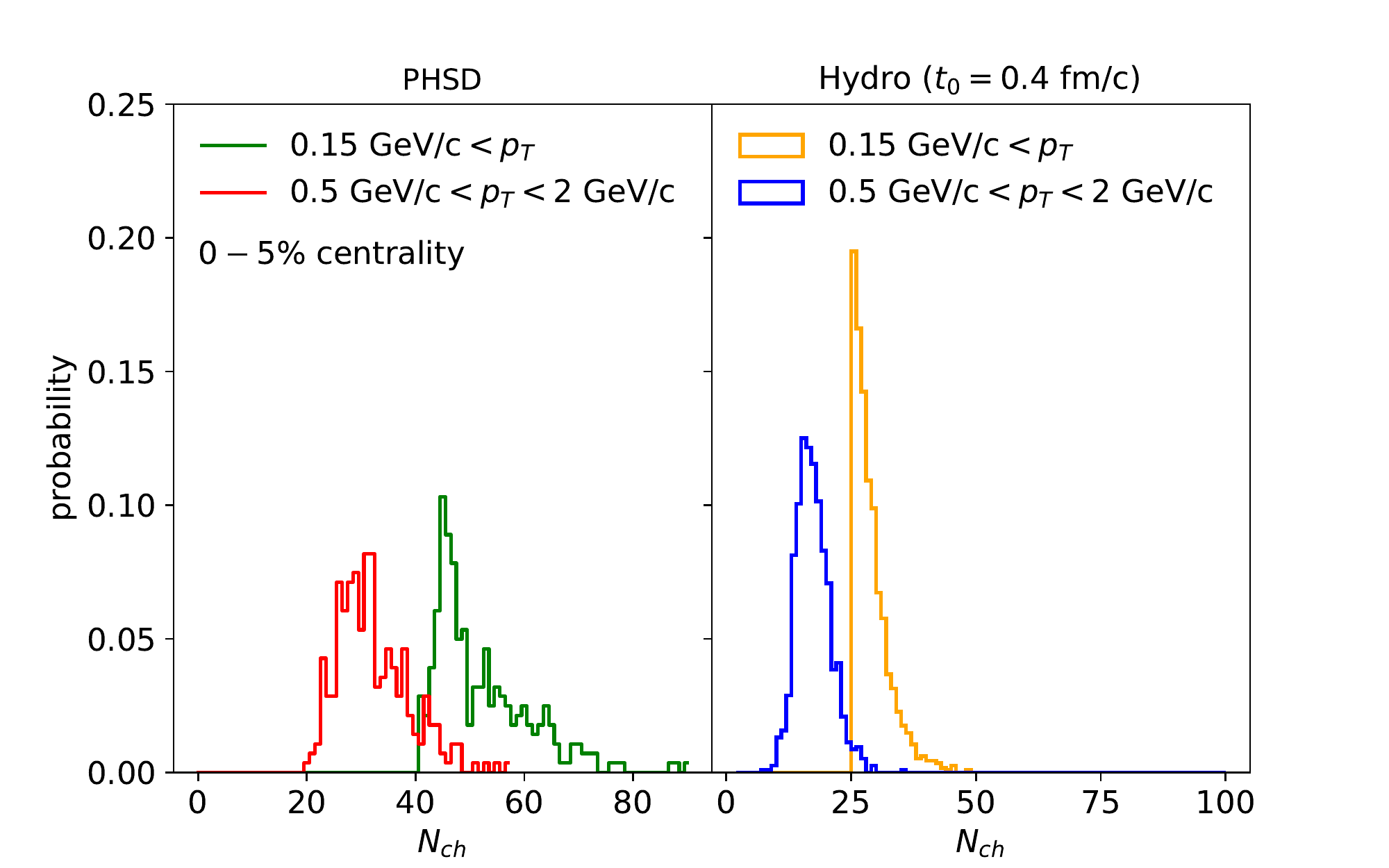}
\put(75,10){\textbf{(b)}}
\end{overpic}
\caption{Event distribution as a function of charged particles for PHSD (left panels) and hydrodynamics (right panels) considering different $p_T$ ranges for minimum bias (a) and 5\% central (b) events.}
\label{fig:nch_hydro_phsd_pTcut}
\end{figure}

Since the difference in the charged particle spectra between PHSD and hydrodynamics is more pronounced at higher and very low $p_T$, we show in Fig.~\ref{fig:nch_hydro_phsd_pTcut} the effect of different cuts in transverse momentum when computing event distribution (normalized to 1) as a function of the number of charged particles for both minimum bias (a) and high-multiplicity (b) events. Left panels correspond to PHSD simulations, while hydrodynamic results with $t_0=0.4$ fm$/c$ are shown in the right panels.
We see, as expected from Figs.~\ref{fig:dNch_dpt} and \ref{fig:nch_hydro_phsd}, that the event distribution of both approaches is different for the different $p_T$ cuts applied in the computation of $N_{ch}$.
While it is expected that the distribution of charged particles changes for different $p_T$ intervals, we will see that it is not the case for the spherocity distribution of charged particles.

\subsection{Transverse spherocity}

The transverse spherocity, $S_{0}$, is defined for a unit vector $\mathbf{ \hat{\rm \mathbf{n}}_{\rm \mathbf{s}} }$ which minimizes the ratio:
\begin{equation}
S_{\rm 0} \equiv \frac{\pi^2}{4}  \underset{\bf\hat{n}_s}{\text{min}} \left( \frac{\sum_{i}|{\bf p_{Ti}} \times {\bf\hat{n}_s}|}{\sum_{i}p_{Ti}} \right)^{2},
\end{equation}
where the sum runs over all charged particles\footnote{As mentioned in Ref.~\cite{Banfi:2010xy}, the minimisation procedure is numerically simplified by the observation that the ${\bf\hat{n}_s}$ that provides the minimal sum always coincides with the transverse direction of one of the ${\bf p_{Ti}}$.} in a chosen pseudorapidity region, that in our case is taken to be $|\eta|<0.5$.
We consider all particles with transverse momentum $p_T>0.15$\,GeV/$c$, analogously to what is done in experiments. Moreover, at least three tracks are required within those $\eta$ and $p_T$ ranges in order to achieve a good spherocity resolution.
The normalization constant $\pi^2/4$ ensures that $S_0$ runs from 0 to 1.
The two limits correspond to event topologies with particular configurations:
\begin{itemize}
\item $S_{0}\rightarrow0$: ``jetty'' events, where all transverse momentum vectors are (anti)parallel or their sum is dominated by a single track;
\item $S_{0}\rightarrow1$: ``isotropic'' events, where the transverse momentum vectors are isotropically distributed. 
\end{itemize}
Especially in p-p collisions \cite{Acharya:2019mzb}, the jetty events are often the result of hard processes while the isotropic one are events in which the soft processes dominate.
\\
Throughout the paper we will refer to transverse spherocity simply as spherocity.

\begin{figure}
\centering
\begin{overpic}[abs,unit=1mm,width=\columnwidth,trim={0 0 35 20},clip]{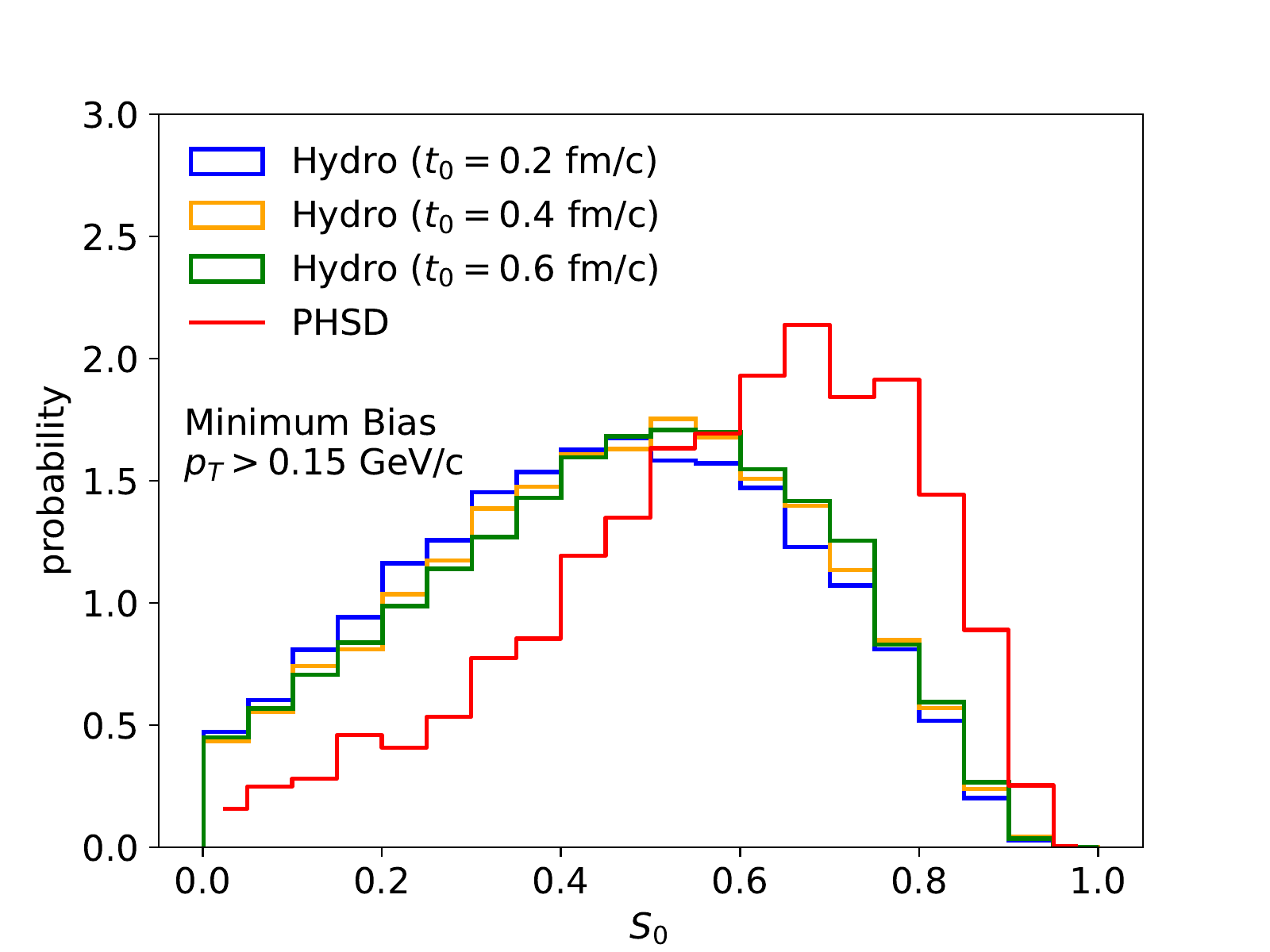}
\put(75,55){\textbf{(a)}}
\end{overpic}
%\par\bigskip % force a bit of vertical whitespace
\begin{overpic}[abs,unit=1mm,width=\columnwidth,trim={0 0 35 20},clip]{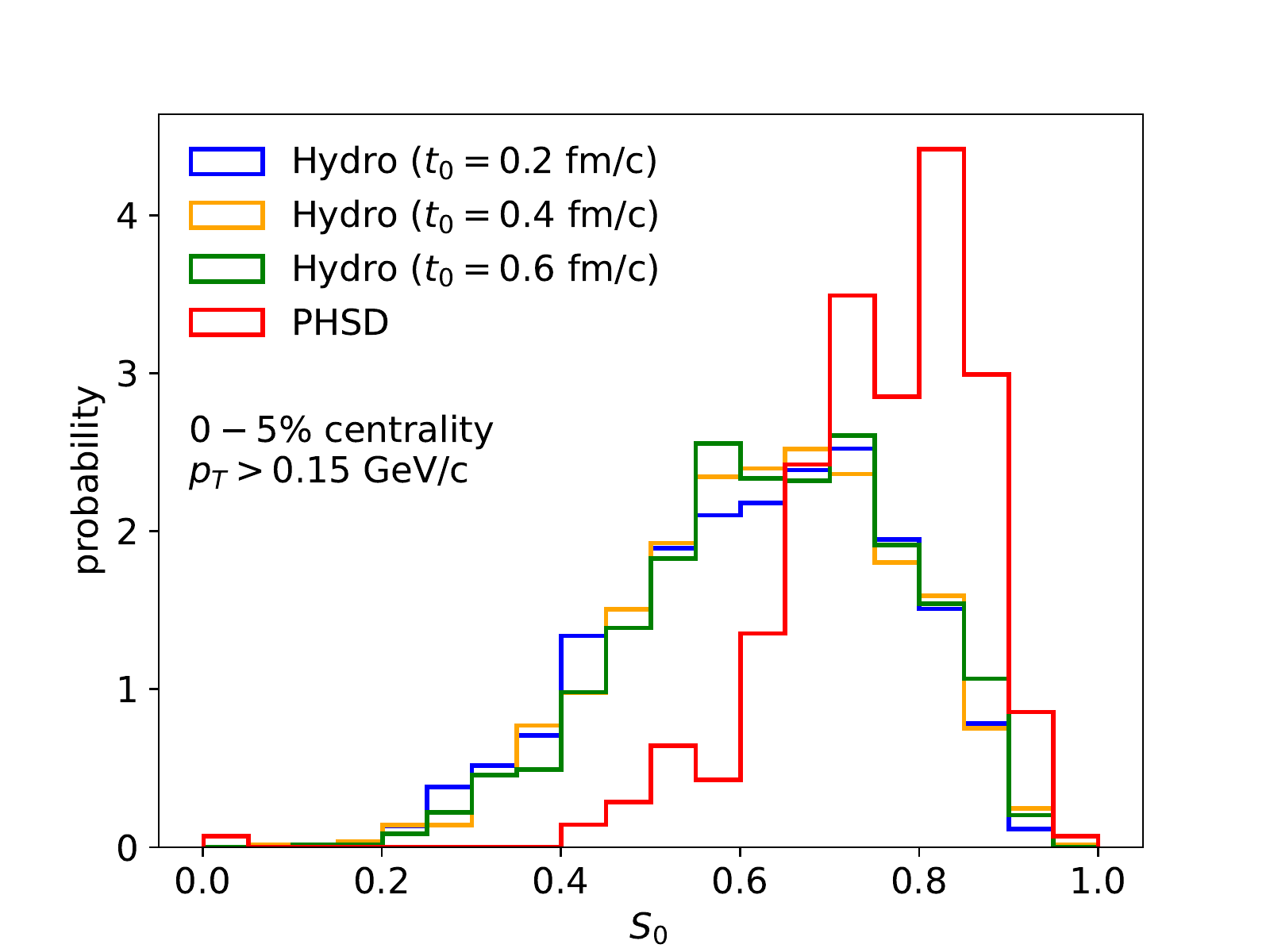}
\put(75,55){\textbf{(b)}}
\end{overpic}
\caption{Spherocity distribution for PHSD events (red line) and hydrodynamics starting from PHSD initial conditions at three different times (green, orange and blue lines). 
Panel (a) corresponds to minimum bias events, while panel (b) shows 5\% central events (with hydrodynamics starting from minimum bias PHSD events and then the 5\% most central events are selected).}
\label{fig:sph_hydro_phsd}
\end{figure}

The distribution of events (normalized to 1) as a function of spherocity is presented in Fig.~\ref{fig:sph_hydro_phsd} for both PHSD (red lines) and hydrodynamic approaches (blue, yellow and green lines) in the case of minimum bias (a) and 5\% most central collisions (b). We see that the spherocity distribution in PHSD is shifted more towards 1 compared to hydrodynamics and similar to predictions from other transport models, such as AMPT \cite{Mallick:2020ium}. Therefore, PHSD events are more isotropic. Probably this difference is partially due to the different event probability as a function of the number of charged particles in the two cases, as visible in Fig.~\ref{fig:nch_hydro_phsd}, but this is not the only reason as discussed in the following.
Further studies extended to (3+1)D hydrodynamics may help to better understand the origin of such discrepancy and also its possible relation to flow development in the two approaches.

\begin{figure}
\centering
\begin{overpic}[abs,unit=1mm,width=\columnwidth,trim={30 0 35 20},clip]{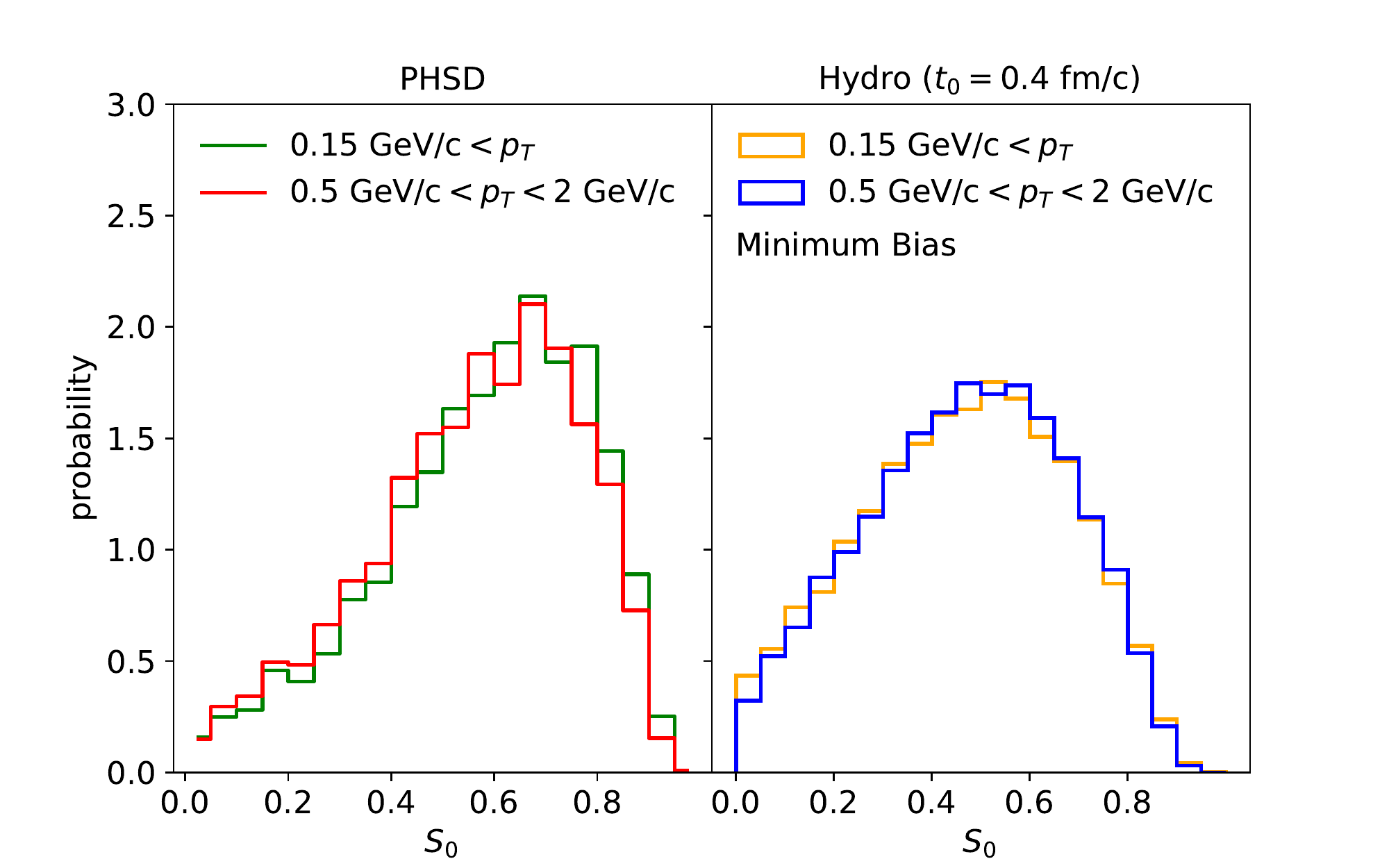}
\put(75,35){\textbf{(a)}}
\end{overpic}
%\par\bigskip % force a bit of vertical whitespace
\begin{overpic}[abs,unit=1mm,width=\columnwidth,trim={30 0 35 20},clip]{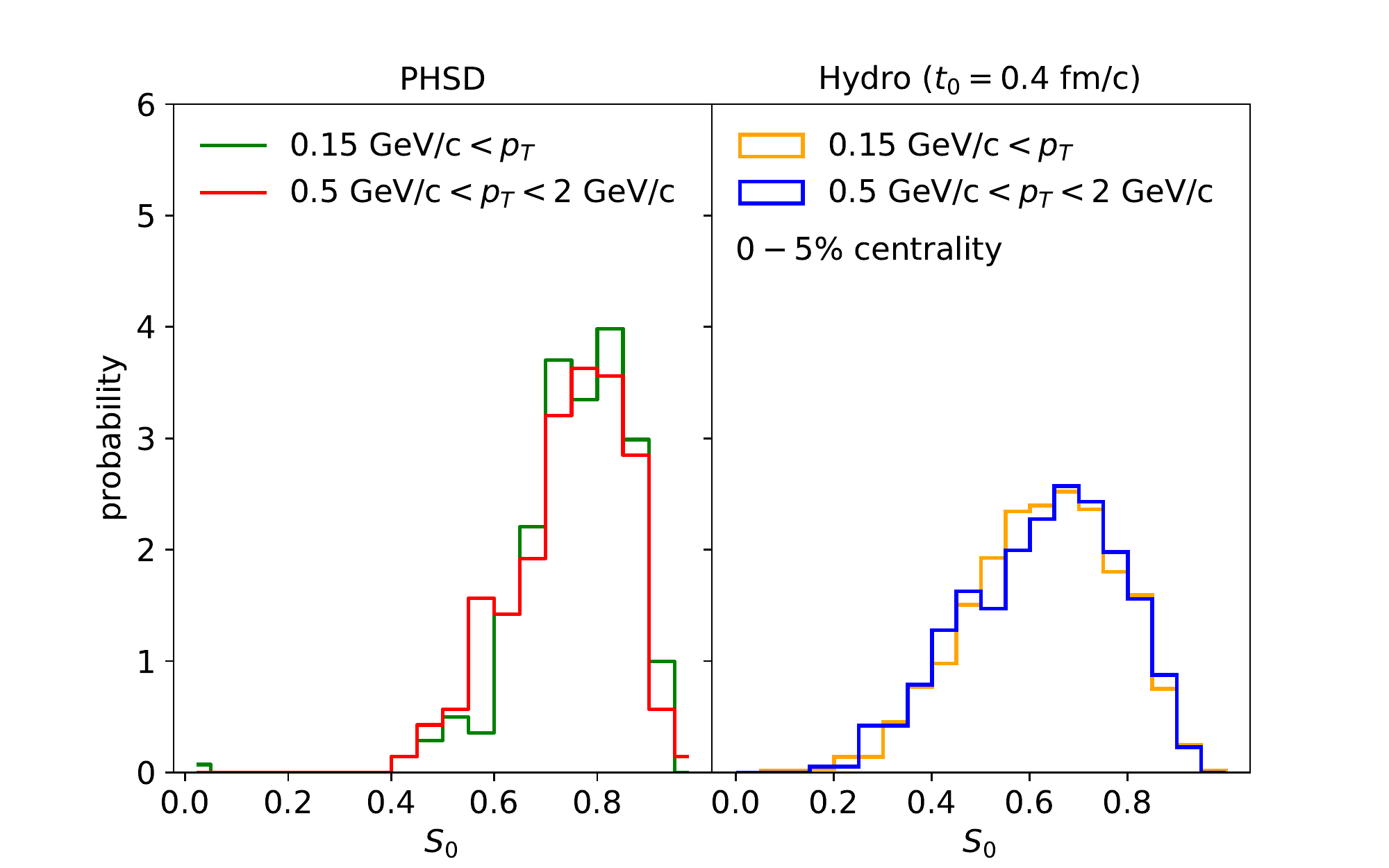}
\put(75,35){\textbf{(b)}}
\end{overpic}
\caption{Event distribution as a function of spherocity for PHSD (left panels) and hydrodynamics (right panels) considering different $p_T$ ranges for minimum bias (a) and 5\% central (b) events.}
\label{fig:sph_hydro_phsd_pTcut}
\end{figure}

As pointed out in Ref.~\cite{Prasad:2021bdq}, event topology is decided by the underlying particle production dynamics and medium effects.
In order to check its connection to the particle multiplicity we plot in Fig.~\ref{fig:sph_hydro_phsd_pTcut} the event distribution (normalized to 1) as a function of spherocity considering the two different transverse momentum cuts applied in Fig.~\ref{fig:nch_hydro_phsd_pTcut} for the charged particle multiplicity, i.e. the spherocity is determined considering only charged particles with $p_T$ in the considered interval.
We notice from Fig.~\ref{fig:sph_hydro_phsd_pTcut} that, even though the event distribution as a function of $N_{ch}$ is substantially modified in both PHSD and hydrodynamics if different $p_T$ cuts are applied in the determination of charged particles (see Fig.~\ref{fig:nch_hydro_phsd_pTcut}), the spherocity distribution does not change within the same approach. This indicates that the disagreement between PHSD and hydrodynamics in the spherocity distribution is not strongly related to the dissimilarity in the charged particle production, but is rather connected to the different description within the two frameworks of the medium produced in small colliding systems.

This finding supports the idea that multi-differential measurements, such as those based on event classification according to multiplicity and event shape, are important tools to study properties of the medium produced in ultra-relativistic proton-nucleus collisions. Performing an event-shape analysis with spherocity in small systems adds to the collection of such tools that have already been discussed for p+p collisions in both theoretical \cite{Khuntia:2018qox} and experimental studies \cite{Acharya:2019mzb, Nassirpour:2020owz}.

\section{Conclusions} 
\label{sec:sum}

We have investigated the evolution of the medium produced in small colliding systems as well as the effects of its far-from-equilibrium dynamics within a microscopic transport and a macroscopic hydrodynamic description.
The initial conditions for p+Pb collisions at LHC energy of $\sqrt{s_{NN}}=5.02$ TeV have been extracted through a Landau-matching procedure from the PHSD transport approach, which describes the full space-time evolution of the relativistic proton-nucleus collision from the initial hard scatterings.
The ensuing medium evolution is followed in PHSD, which includes inherently nonequilibrium effects, and by means of the (2+1)D viscous hydrodynamic model VISHNew, which takes into account the large deviations from local equilibrium expected in a proton-nucleus collision.
Different initial times have been considered for the start of the hydrodynamic simulations and we have compared in the two approaches quantities like the energy density, the flow velocity, the bulk viscous pressure and the spatial and momentum eccentricities.
In PHSD the energy density, while rapidly decreases as the medium expands, is highly inhomogeneous in the transverse plane during the whole evolution; in the hydrodynamic simulations the initial hot spots dissolve more efficiently than in PHSD, but still the energy density profile keeps a high degree of inhomogeneity due to the smaller size of the medium produced in p+Pb collisions with respect to heavy-ion reactions.
Moreover, for later initialization times the system is already more diluted and has less time to smooth the hot spots in the energy density that remain more visible throughout the evolution.
The spatial irregularity of the energy density are quantified in terms of its Fourier modes: their strength is similar in PHSD and VISHnew at the initial times but after the first fm$/c$ the values of shorter wavelength modes rapidly decrease with respect to the zero mode in the hydrodynamic medium, while a high degree of inhomogeneity is maintained in the microscopically evolving medium.
The evolution of the bulk viscous pressure in small systems is quicker than that in heavy-ion collisions. The bulk viscous pressure is initially very large and negative due to the large initial expansion rate. Then its magnitude in PHSD experiences a power-law decay but remains nonzero during the collision evolution, while in the hydrodynamic case it approaches quickly zero within about half fm$/c$. Moreover, we found that in hydrodynamics the bulk viscous pressure loses quickly memory of the initial conditions extracted from PHSD at various starting times (hence quite different), relaxing to a single trajectory with a behaviour resembling that of hydrodynamic attractors.

The gained understanding of the nonequilibrium effect in the evolution of bulk QCD matter in proton-nucleus collisions may help to identify the impact of these effects on final observables, such as strangeness enhancement and collective flow that are strictly linked to the QGP formation. In this respect it is becoming more and more important the event-shape determination, that can be used along with centrality selection for a multi-differential categorization of the events in which one can study more efficiently the QGP signatures.
We have performed a transverse spherocity analysis in p+Pb collisions at LHC energy with PHSD and hydrodynamic simulations starting from the same initial conditions in order to identify the effects coming from the different medium descriptions underlying the two approaches. We found that the spherocity distribution in PHSD is more towards 1 with respect to the hydrodynamic result and similar to prediction from other transport models. Therefore, the PHSD evolution favors more isotropic event topologies. Even though this dissimilarity is partially due to the different final charged particle production, it mainly comes from the different description within the two frameworks of the medium produced in small colliding systems. This finding supports the idea that multi-differential measurements, such as those based on event classification according to multiplicity and spherocity, are useful to study observables of the medium produced in ultrarelativistic proton-nucleus collisions.

\section*{Acknowledgements}
The authors acknowledge inspiring discussions with  W. Cassing, J.-F. Paquet, T. Song, V. Voronyuk.
L.O. and E.B. acknowledge support by the Deutsche Forschungsgemeinschaft (DFG, German Research Foundation) through the grant CRC-TR 211 "Strong-interaction matter under extreme conditions" - project number 315477589 - TRR 211 and by the German Academic Exchange Service (DAAD).
L.O. has been in part financially funded by the Alexander von Humboldt-Stiftung.
W.F., P.M. and S.A.B. has been supported by the U.S. Department of Energy Grant no. DE-FG02-05ER41367.
The computational resources have been provided by the Goethe-HLR Center for Scientific Computing.

\bibliography{References}

\end{document}